\input harvmac
\overfullrule=0pt
\input amssym
\font\small=cmr7      
\def\Zz{Z}
\def\Zzt{\tilde Z}
\def\gg{g}
\def\ggt{\tilde g}     
\def\Aa{\A}
\def\Aat{\tilde\A}    
\def\hh{h}
\def\hht{\tilde h}            
\def\gam{\gamma}
\def\gamt{\tilde\gamma}                                                                                                                                          
\def\bx{{\bf x}}

\def\nox{{\scriptstyle{\times \atop \times}}}

\def\Zop{{\Bbb Z}}
\def\Rop{{\Bbb R}}
\def\Cop{{\Bbb C}}
\def\Pop{{\Bbb P}}

\def\boxit#1{\vbox{\hrule\hbox{\vrule\kern5pt
vbox{\kern5pt#1\kern5pt}\kern5pt\vrule}\hrule}}
\def\vac{|0\rangle}
\def\bvac{\langle 0|}

\def\q{\hbox{\rm q}}
\def\A{{\cal A}}
\def\J{{\cal J}}

\def\H{{\cal H}}

\def\N{{\cal N}}
\def\P{{\cal P}}

\def\im{\hbox{Im}}
\def\re{\hbox{Re }}

\def\zb{{\bar z}}
\def\tx{\tilde\xi}
\def\half{{\scriptstyle {1\over 2}}}
\def\da{{\dot a}}
\def\db{{\dot b}}

\def\bsigma{{\bf \sigma}}
\def\ker{\hbox{Ker }}
\def\floop{{\hbox{\small loop}}}
\def\tree{{\hbox{\small tree}}}
\def\tr{\hbox{tr}}

\parindent=0pt
\parskip=6pt

\def\vs{\vskip6pt}

\def\sqr#1#2{{\vbox{\hrule height.#2pt\hbox{\vrule width
.#2pt height#1pt \kern#1pt\vrule width.#2pt}\hrule height.#2pt}}}

\def\half{{\scriptstyle{1\over 2}}}

\def\rhoP{\rho_\Pi}

\Title{\vbox{\baselineskip12pt
\hbox{hep-th/0703054}}}
{\vbox{\centerline{Tree and Loop Amplitudes}
\bigskip
\centerline{in Open Twistor String Theory}}}
\smallskip
\centerline{Louise Dolan
}
\smallskip
\centerline{\it Department of Physics}
\centerline{\it
University of North Carolina, Chapel Hill, NC 27599-3255}
\bigskip
\smallskip
\centerline{Peter Goddard}
\smallskip
\centerline{\it Institute for Advanced Study}
\centerline{\it Olden Lane, Princeton, NJ 08540, USA}
\bigskip
\bigskip

We compute the one-loop gluon amplitude for the open twistor
string model of Berkovits, using a symmetric form of the vertex operators.
We discuss the classical solutions in various topologies and instanton sectors
and the canonical quantization of the world sheet Lagrangian. We derive the
$N$-point functions for the gluon tree and one-loop amplitudes,
and calculate a general one-loop expression for the current algebra.

\Date{}

\nref\Wittenone{E. Witten, ``Perturbative Gauge Theory A String Theory
In Twistor Space'', Commun.\ Math.\ Phys.\  {\bf 252} (2004) 189;
hep-th/0312171.}

\nref\Berkovitsone{N. Berkovits, ``An Alternative String Theory in
Twistor Space for N = 4 Super-Yang-Mills'',
Phys.\ Rev.\ Lett.\  {\bf 93}  (2004) 011601; hep-th/0402045.}

\nref\BeWi{E. Witten and N. Berkovits, ``Conformal Supergravity in
Twistor-String Theory'',
JHEP {\bf 0408}, 009 (2004); hep-th/0406051.}

\nref\BeMo{N. Berkovits and L. Motl, ``Cubic Twistorial String Field Theory'',
JHEP {\bf 0404} (2004) 056; hep-th/0403187.}

\nref\BSS{L. Brink, J. Scherk, and J. Schwarz, ``Supersymmetric Yang Mills
Theories'', Nucl. Phys. {\bf B121} (1977) 77.}

\nref\BG{F. Berends and W. Giele, ``Recursive Calculations for Processes
with N Gluons", Nucl. Phys. {\bf B306} (1988) 759.}

\nref\CSWone{F. Cachazo, P. Svrcek and E. Witten, 
``MHV Vertices and Tree Amplitudes in Gauge Theory'',
JHEP {\bf 0409}, (2004) 006;
hep-th/0403047.}

\nref\BDIP{N. Bjerrum-Bohr, D. Dunbar, H. Ita, and W. Perkins,
``MHV-vertices for Gravity Amplitudes'', JHEP 0601 (2006)009; hep-th/0509016.}

\nref\BGK{F. Berends, W. Giele, and H. Kuijf,
``On Relations Between Multi-gluon and Multi-graviton Scattering'',
Phys. Lett. {\bf B211} (1988) 91.}

\nref\GSB{M. Green, J. Schwarz, and L. Brink, ``$N=4$ Yang-Mills and $N=8$
Supergravity as Limits of String Theories'', Nucl.\ Phys.\ B {\bf 198} (1982) 474.}

\nref\BDK{Z.~Bern, L.~J.~Dixon and D.~A.~Kosower,
``Dimensionally Regulated Pentagon Integrals'',
Nucl.\ Phys.\ B {\bf 412} (1994) 751;
hep-ph/9306240.}

\nref\Forde{D.Forde and D.Kosower, ``All-multiplicity One-loop Corrections
to MHV amplitudes in QCD'', Phys.\ Rev.\ D {\bf 73} (2006) 061701;
hep-ph/0509358.}

\nref\Dixonone{
C.Berger, Z.Bern, L.Dixon, D.Forde and D.Kosower,
``Bootstrapping One-loop QCD Amplitudes with General Helicities'',
Phys.\ Rev.\ D {\bf 74} (2006) 036009; hep-ph/0604195.}

\nref\Dixontwo{
C.Berger, Z.Bern, L.Dixon, D.Forde and D.Kosower,
``All One-loop Maximally Helicity Violating Gluonic Amplitudes in QCD'',
hep-ph/0607014.}

\nref\CSW{F. Cachazo, P. Svrcek and E. Witten,
``Twistor Space Structure Of One-loop Amplitudes In Gauge Theory'',
JHEP {\bf 0410} (2004) 074; hep-th/0406177.}

\nref\BBKR{I.~Bena, Z.~Bern, D.~A.~Kosower and R.~Roiban,
``Loops in Twistor Space'', Phys.\ Rev.\ D {\bf 71} (2005) 106010,
hep-th/0410054.}

\nref\PT{S. Parke and T. Taylor, ``An Amplitude for N Gluon Scattering'',
Phys. Rev. Lett. {\bf 56} (1986) 2459.}

\nref\MaPa{M. Mangano and S. Parke, ``Multiparton Amplitudes In Gauge
Theories'',  Phys.\ Rept.\  {\bf 200} (1991) 301.}

\nref\DG{L. Dolan and P. Goddard, to appear}

\nref\RN{H. Romer and P. van Nieuwenhuizen, ``Axial Anomalies in $N=4$
Conformal Supergravity'', Phys. Lett. {\bf B 162} (1985), 290.}

\nref\NB{
N. Berkovits, ``Calculation of Green-Schwarz Superstring Amplitudes Using the $N=2$ Twistor-String
Formalism'',   Nucl. Phys. B {\bf 395}, 77 (1993); hep-th/9208035.}
~                                                                    

\nref\FZ{I.B. Frenkel and Y. Zhu, ``Vertex Operator Algebras Associated to Representations of
Affine and Virasoro Algebras'', Duke Math J. {\bf 66}, 123 (1992).} 

\nref\FO{M.D. Freeman and D. Olive,
``BRS Cohomology in String Theory and the No Ghost Theorem'',
Physics Lett. B175 (1986) 151; and
``The Calculation of Planar One Loop Diagrams in String Theory Using the BRS
Formalism'', B175 (1986) 155.}

\nref\Zhu{Y. Zhu, ``Modular Invariance of Characters of Vertex Operator
Algebras'', J. Amer. Math. Soc. {\bf 9} (1996) 237-302.}

\nref\MT{G. Mason and M. Tuite, ``Torus Chiral n-Point Functions for Free
Boson and Lattice Vertex Operator Algebras'', Comm. Math. Phys. {\bf 235} (2003)
47-68.}

\nref\Weyl{Hermann Weyl,   ``The Classical Groups, Their Invariants and Representations", Princeton Press (1977) 279.}

\nref\Mumford{David Mumford, {\it Tata Lectures on Theta I}, Progress in
Mathematics, vol. 28, Birkhauser: Boston, 1983.}

\newsec{Introduction}

Twistor string theory \refs{\Wittenone, \Berkovitsone, \BeWi}
offers an approach to formulating a QCD string. 
Unlike conventional string theory, the twistor string has a finite number of states. These include massless ones described by $N=4$  super Yang Mills theory 
coupled to $N=4$ conformal supergravity in four spacetime dimensions. 
The original theory \refs{\Wittenone} is a topological
string theory, where a finite number of states arises in the usual way.
The alternative formulation \refs{\Berkovitsone, \BeWi} has both Yang-Mills
and supergravity particles occurring in an open string sector. Here
the absence of a Regge tower of states is due to the absence of 
operators involving momentum for massive particles. 
Since the twistor string theory is some N=4 Yang-Mills field theory coupled to N=4 conformal supergravity and possibly
a finite number of closed string 
states, we infer that this field theory system is ultraviolet finite.

In \refs{\Wittenone}, gluon amplitudes with $\ell$ loops, 
and with $d+1-\ell$ negative helicity gluons and
the rest positive helicity gluons, were associated with a topological string
theory with D-instanton contributions of degree $d$. Beyond tree level,
the occurrence of conformal supergravity states is believed to modify
the gluon amplitudes from Yang Mills theory. 
The other version of twistor string theory \refs{\Berkovitsone}
uses a set of first order `b,c' world sheet variables 
with open string boundary conditions, and world sheet instantons.
The target space of both models is the supersymmetric version of twistor space, $\Cop\Pop^{3|4}$.
In \refs{\BeMo}, a path integral construction of the tree amplitudes
outlined in
\refs{\Berkovitsone}, is used to compute the three gluon tree amplitude.
Some $n$-point Yang-Mills 
and conformal supergraviton string tree amplitudes are derived in {\BeWi},
which discusses both models and general features of loop amplitudes. 
Some earlier computations of conventional field theory amplitudes in a helicity basis can be found in
\refs{\BSS-\CSWone} for Yang Mills trees, \refs{\BDIP,\BGK} for gravity trees, 
and \refs{\GSB -\Dixontwo} for
the Yang Mills loop. General features of the twistor structure of the Yang Mills loop are described in \refs{\CSW,\BBKR}.

In this paper, we calculate an expression for the one-loop n-gluon amplitude in
Berkovits' open twistor string theory. In section 2, we review the
world sheet action, establishing a convenient notation and selecting  
 a gauge where the 
world sheet abelian gauge fields have been gauged to zero. Then the topology,
or  instanton number,
resides in the boundary conditions, for an open string, or the transition functions relating different gauge patches on the world sheet, for a closed string. We discuss the classical solutions on the sphere, disk, torus and cylinder.

In section 3, we discuss the canonical quantization of the string,  gauge invariance in the corresponding operator formalism and the construction of vertex operators corresponding to gluons.

In section 4, the $n$-point gluon open string tree amplitudes are calculated, and they match the Parke-Taylor form \refs{\PT,\MaPa}, up to
double trace terms from the current algebra. This extends the
three-point amplitudes found via a path integral in \refs{\BeMo},
and provides an explicit
oscillator quantization of some of the tree amplitudes found in \refs{\BeWi}.

In section 5, the gluon open string one-loop amplitudes
are described as a product of contributions from twistor fields, ghost fields,
and the current algebra. We compute the n-gluon one-loop twistor field
amplitude. Both the $n$-point tree and loop amplitudes are computed for
the maximally helicity violating (MHV) amplitudes, where the instanton number
takes values $1,2$. 
This can be extended straightforwardly to any instanton number, leading to amplitudes with
arbitrary numbers of negative and positive helicity gluons. 

In section 6, the general expressions for two-, three- and four-point one-loop current algebra correlators are given for an
arbitrary Lie group. They are given in terms of  Weierstrass $\P$ and $\zeta$ functions. We expect to discuss the derivation of these expressions and their 
generalizations, using recursion relations, in a later paper \refs{\DG}.

In section 7, we combine the parts and construct the 
four-point MHV one-loop gluon amplitude of the open twistor string.
We show how the delta function vertex operators of the twistor string
lead to a form of the final integral that is a simple product of the
current algebra loop and the twistor fields loop. We do not discuss here the infrared regularization of the loop amplitude, nor how the gauge group of the current algebra
is ultimately determined \refs{\RN}.

\vfill\eject
\newsec{The Action and Classical Solutions}

{\sl Equations of Motion and Boundary Conditions.} The world sheet action for the twistor string introduced by Berkovits can be written in the form
\eqn\act{S = S_{YZ} + S_{\rm ghost} + S_G} where $S_G$ represents a 
conformal field theory with $c=28$ and $S_{YZ}$ is given by
\eqn\cov{{S_{YZ}}
= \int i\left [Y^{I\mu}D_\mu Z_{IS}+Y^I_\mu\epsilon^{\mu\nu}D_\nu
Z_{IP}\right ] g^\half d^2x\,,}
with $D_\mu = \partial_\mu -i A_\mu$ and $1\le I\le 8$, and the fields $Z_S, Z_P$ and $Y_\mu$ are homogeneous coordinates in the complex projective twistor superspace $\Cop\Pop^{3|4}$ and are world-sheet scalars, pseudo-scalars and vectors, respectively.
For the action to be real, the fields must satisfy the conditions $\overline{Y^{I\mu}}= -Y^{I\mu}$ for the bosonic components ($1\leq I\leq 4$) and $\overline{Y^{I\mu}}= Y^{I\mu}$ for the fermionic components ($5\leq I\leq 8$), $\overline{Z_{IS}}=Z_{IS}$, 
$\overline{Z_{IP}}=Z_{IP}$, but we must also have
$\overline{A_\mu}=-A_\mu$, {\it i.e.} $A_\mu$ has to be pure imaginary.

The action \act\ gives rise to  the equations of motion
\eqn\eom{D_\mu Z_S+ {\epsilon_\mu}^\nu D_\nu Z_P=0,\qquad D'_\mu Y^\mu=0, 
\qquad D'_\mu\epsilon^{\mu\nu}Y_\nu=0}
where $D'_\mu=(\partial_\mu+iA_\mu)g^\half$,
to the constraint,
\eqn\con{Y^\mu Z_S + Y_\nu\epsilon^{\nu\mu}Z_P=0\;,}
and to the  end condition on the open string,
\eqn\endc{Y^\mu n_\mu \delta Z_S + Y_\mu\epsilon^{\mu\nu}n_\nu\delta Z_P=0\,,}
where $n^\nu$ is a vector normal to the boundary.
The end condition \endc\ will be satisfied if 
\eqn\bound{Y_\mu n^\mu\cos\alpha+Y_\mu\epsilon^{\mu\nu}n_\nu\sin\alpha=0\,,
\qquad Z_S\sin\alpha-Z_P\cos\alpha=0\,,}
for some function  $\alpha$, varying over the boundary, and continuous up to multiples of $\pi$.  The function $\alpha$ changes under
gauge transformations, which we shall now discuss.

In the case of a Euclidean signature for the world sheet, 
we can write 
\eqn\euc{Y^{\mu}D_\mu Z_{S}+Y_\mu\epsilon^{\mu\nu}D_\nu
Z_{P}=Y^zD_z \Zzt+Y^\zb D_\zb \Zz\;,}
where $z=x_1+ix_2$, $\zb =x_1-ix_2$, $\Zz = Z_S-iZ_P$, $\Zzt=Z_S+iZ_P$, 
$D_z=\partial_z - i A_z$, $A_z=\half (A_1-iA_2)$, {\it etc}.
With this notation, the equations of motion become 
\eqn\eoma{D_\zb \Zz=D_z \Zzt=0,\qquad D_z'Y^z=D'_\zb Y^\zb=0,}
together with the constraints
$$ Y^\zb \Zz=Y^z\Zzt=0,$$
the boundary conditions
\eqn\endca{\Zzt=U\Zz,\qquad Y^zn_z=-U^{-1}Y^\zb n_\zb,}
where $U=e^{2i\alpha}$ in terms of \bound, and reality conditions $\bar Z=\tilde Z$, $\overline{Y^z} = - Y^{\bar z}$ for bosonic components, and $\overline{Y^z} = Y^{\bar z}$ for fermionic  components, $\overline{A_z}=-A_\zb$. The reality conditions imply that on the boundary $|U|=1$.

{\sl Gauge Invariance.} The action has  two abelian gauge invariances,
\eqn\gaugea{Y^\zb\mapsto \gg^{-1}Y^\zb,\quad 
\Zz\mapsto \gg \Zz, \quad A_\zb\mapsto A_\zb -i\gg^{-1}\partial_\zb \gg\,,}
\eqn\gaugeb{Y^z\mapsto  \ggt^{-1}Y^z, \quad \Zzt\mapsto \ggt \Zzt,\quad A_z\mapsto A_z
-i \ggt^{-1}\partial_z \ggt\,,}
where  $\gg=e^{\psi+i\varphi}, \,\, \ggt=e^{-\psi+i\varphi}$, each in $GL(1,\Cop)$, so that 
\eqn\gaugec{A_\mu\mapsto A_\mu+\partial_\mu\varphi +{\epsilon_\mu}^\nu\partial_\nu\psi\,,}
and $\varphi,\psi$ need to be pure imaginary, {\it i.e.} $\bar \gg =\ggt$ (reducing the gauge group to one copy of $GL(1,\Cop)$), if the reality condition on $A_\mu$ is to be maintained.
$A_\zb, A_z,$ can be thought of as components, $\Aa_\zb,\Aat_z,$  taken from different gauge potentials, $\Aa_\mu, \Aat_\mu,$ associated with the transformations $\gg, \ggt $, respectively. 

The gauge invariance of the theory can be used in general to set the potential $A_\mu=0$, with the vestige of the gauge structure residing in the boundary conditions or the gauge transformations which relate the fields on different patches, in the case of world sheets with non-trivial global topology (and in the components $\Aa_z,\Aat_\zb,$ that do not appear explicitly in the action). In a gauge with $A_z=A_\zb=0$, the equations of motion for $\Zz,\Zzt$ become $\partial_\zb \Zz=\partial_z\Zzt=0$, {\it i.e.} $\Zz\equiv \Zz(z), \Zzt\equiv \Zzt(\zb)$.

{\sl Solutions on the Sphere.} If the world sheet is the sphere $S^2$, corresponding to closed string boundary conditions, mapped stereographically onto the plane, the potentials are defined by functions on two patches, $A^>_\mu$ on $S^>_\mu=\{z: |z|>1-\epsilon\}$ and $A^<_\mu$ on $S^<_\mu=\{z: |z|<1+\epsilon\}$, for some $\epsilon >0$, with
\eqn\patch{A^>_\zb-A^<_\zb=-i\gg^{-1}\partial_\zb \gg,\qquad A^>_z-A^<_z=-i\ggt^{-1}\partial_z \ggt\qquad
\hbox{for }1+\epsilon >|z|>1-\epsilon.}
We can apply gauge transformations $\gam^>, \gamt^>, \gam^<, \gamt^<$, on the two patches separately which map all of $A^>_\zb,A^>_z, A^<_\zb,A^<_z$ to zero, so that the gauge transformations relating the patches become
$$
\hh=\gam^>\gg(\gam^<)^{-1}, \quad \hht=\gamt^>\ggt(\gamt^<)^{-1}, \quad\hbox{so that}\quad
\partial_\zb \hh=\partial_z\hht=0
$$
implying $\hh\equiv \hh(z), \hht\equiv \hht(\zb)$. Maintaining the reality conditions implies that $\bar \hh=\hht$. As $z$ encircles the unit circle $|z|=1$, the phase of $\hh(z)$ will increase by $-2\pi n$
for some integer $n$, so that $\log(z^n\hh(z))$ is single-valued in the annulus $1+\epsilon >|z|>1-\epsilon$. By writing $\log(z^n\hh(z))$ as the sum of two functions, one regular at the origin and the other at infinity, we can obtain gauge transformations on the two patches that will maintain $A_\zb^>= A_\zb^<=0$ while replacing the transition function $\hh(z)$ by $z^{-n}$. Correspondingly, $\hht(\zb)$ is replaced by $\zb^{-n}$. Thus, for the sphere, we can always choose a gauge in which the components of the potential occurring in the equations of motion are zero and the $\Zz,\Zzt$ fields on the two patches are related by
\eqn\Zrel{
\Zz^>(z)=z^{-n}\Zz^<(z),\qquad \Zzt^>(\zb)=\zb^{-n}\Zzt^<(\zb).
}
(See Appendix A.) These conditions have a solution provided that $n\geq 0$, in which case $\Zz^<(z), \Zzt^<(\zb)$ are polynomials of order $n$ which are complex conjugates of one another. 

{\sl Solutions on the Disk.} If the world sheet is the disk, $D^2=\{z:|z|\leq 1\}$, appropriate to open string tree amplitudes, we can again choose a gauge in which $A_z=A_\zb=0$, so that $\Zz$ and $\Zzt$ are analytic functions of $z$ and $\zb$, respectively, and the residue of the gauge structure is only left in the boundary conditions \endca. If the phase of $U(z)$ in this equation changes by $-2\pi n$, $n$ an integer, as $z$ goes round the unit circle, 
we can write $\log(z^n U(z))=f_<(z)+f_>(z)$ on the unit circle, where $f_>(z), f_<(z)$ are defined and holomorphic in $|z|\geq 1$ and $|z|\leq 1$, respectively. The fact that $|U|=1$ on the unit circle implies that $\overline{f_<(z)}=-f_>(1/\zb)$. If we now apply the gauge transformation $\gamma=e^{-f_>(1/\zb)}, \tilde\gamma=e^{f_<(z)}$, $U\mapsto
\gamma U\tilde\gamma^{-1}=z^{-n}$ on $|z|=1$. The boundary condition only has non-trivial solutions for $n\geq 0$ and the general solution satisfying the reality and boundary conditions is then
\eqn\dgs{\Zz(z)=\sum_{m=0}^n \Zz_m z^m,\qquad \Zzt(\zb)=\sum_{m=0}^n \bar\Zz_m \zb^m,}
where $\Zz_m=\bar \Zz_{n-m}$. 

{\sl Solutions on the Torus.} If the world sheet is a torus, $T^2$, corresponding to a closed string loop amplitude, we can describe it by identifying points of the complex plane related by translations of the form $z\mapsto z+m_1+n_1\tau$, $m_1, n_1\in\Zop$, for a given modulus $\tau\in\Cop$. The various copies of the fundamental region $\{z=x+y\tau: 0\leq x, y <1\}$ have to be related by gauge transformations, $g_a, \tilde g_a$,
\eqn\trans{A_\zb(z+a)=A_\zb(z)-i\gg_a^{-1}\partial_\zb \gg_a, \quad
A_z(z+a)=A_z(z)-i\ggt_a^{-1}\partial_z \ggt_a,}
where $a=m_1+n_1\tau$, $ m_1,n_1\in\Zop.$ The gauge transformations have to satisfy 
\eqn\consist{\gg_{a+b}(z)=\gg_a(z+b)\gg_b(z)=\gg_b(z+a)\gg_b(z),}
and similarly for $\ggt_a$, and the reality condition $\overline{\gg_a(z)}=\ggt_a(\zb)$.

We can apply gauge transformations $\gam, \gamt$ to $\A, \tilde\A$ to set $A_z=A_\zb=0$ over the plane,
changing the gauge transition functions $\gg_a(z)\mapsto \hh_a = \gam(z+a)\gg_a(z)\gam(z)^{-1}$,
$\ggt_a(z)\mapsto \hht_a = \gamt(z+a)\ggt_a(z)\gamt(z)^{-1}$, where $\hh_a,\hht_a$ are holomorphic functions of $z,\zb,$ respectively. In this gauge,
\eqn\transZ{\Zz(z+a)=\hh_a(z)\Zz(z),\qquad \Zzt(\zb+\bar a)=\hht_a(\zb)\Zzt(\zb).}
Writing $ h_a(z)=e^{i \rho_a(z)}$, \consist\ implies
\eqn\top{\rho_1(z+\tau)-\rho_1(z)-\rho_\tau(z+1)+\rho_\tau(z)=-2\pi n,}
for some integer $n$, which describes the topology of the solution. A particular gauge transformation that possesses this property and, more generally, satisfies \consist, is 
\eqn\gastan{\hh_a^0(z) = \exp\left({-\pi n (a-\bar a) \over \im\tau}(z
+ {a\over 2}) + i \pi n m_1n_1 + i\eta_a \right)
}
where $\eta_{a+b}=\eta_a+\eta_b$. 
In fact, if $\hh_a(z)$ satisfies \top, we would need to make a further gauge transformation to bring it into the standard form \gastan.

If we let $\eta_a =\pi m_1\epsilon-\pi n_1\epsilon'$, the translation property of $\Zz$ is
\eqn\transZa{\Zz(z+1)= e^{i\pi\epsilon}\Zz(z),\qquad  
\Zz(z+\tau)=e^{-i\pi(\epsilon'+ n(2z+\tau))}\Zz(z),}
which are the defining relations for an $n$-th order theta function with characteristics $\epsilon,\epsilon'$, and we must have $n>0$ for non-trivial solutions. If the usual theta function is denoted by
\eqn\thetach{
\theta\left[{\epsilon\atop \epsilon'}\right](\nu,\tau)=
\sum_{m\in\Zop}\exp\left\{i\pi(m+\half\epsilon)^2\tau+2\pi i(m+\half\epsilon)\nu+ \pi im\epsilon' +\half\pi i\epsilon\epsilon'\right\},}
the space of $n$-th order theta functions is spanned by the $n$ functions
\eqn\nthorder
{\theta\left[{{1\over n}(\epsilon+2p)\atop \epsilon'}\right](nz,n\tau), \quad p=0, 1, \ldots n-1.}
Thus each of the components of $\Zz$ has an expansion of the form
\eqn\nthordexp{
\Zz^I(z)=\sum_{p=0}^{n-1}c_p^I\,\theta\left[{{1\over n}(\epsilon+2p)\atop \epsilon'}\right](nz,n\tau),}
where $1\leq I\leq 8$.

{\sl Solutions on the Cylinder.} For an open string loop amplitude, we need to consider the world sheet being a cylinder, $C^2$, which we take to be the strip region $\{z: 0\leq \re z\leq \half\}$ with $z$ identified with $z+n\tau$, where $\tau$ is pure 
imaginary. The equations \trans\ and \consist\ hold but with $a, b$ restricted to be integral multiples of $\tau$. The fields $Z(z), \tilde Z(\zb)$ have to satisfy boundary conditions on $\re z=0, \half$,
$$ \Zzt(-iy)= U_0(y)\Zz(iy),\qquad \Zzt(\half-iy)= U_\half(y)\Zz(\half +iy),\qquad |U_0(y)|=|U_\half(y)|=1,$$ for real $y$. 

We can again work in a gauge in which $A_z=A_\zb=0$, where $\Zz,\Zzt$ are analytic functions of $z, \zb,$ respectively. We can find the gauge transformation $\gam_0=e^{i\rho_0}$ to take us to such a gauge by solving the equation $\A_\zb=-\partial_\zb\rho_0$ and we can impose the boundary condition $\rho_0=\half \varphi_0$ on $\re z=0$, where $U_0(y)=e^{i\varphi_0(y)}$ with $\varphi_0(y)\in\Rop$. Under the gauge transformation, $\Zz\mapsto \gam_0 \Zz$, $\Zzt\mapsto \bar\gam_0\Zzt$, so $U_0(y)\mapsto \bar\gam_0 (iy)U_0(y)\gam_0(iy)^{-1}=1$ for real $y$. So we can choose a gauge in which $A_z=A_\zb=0$ and $\Zz,\Zzt$ are analytic functions of $z,\zb,$ respectively, and $\Zz,\Zzt$ are real on $\re z=0$. In this gauge we can extend the definition of the fields into the region 
$\{z: -\half\leq \re z\leq 0\}$ by reflection about $\re z=0$: $ \Zz(z)=\overline{\Zz(-\zb)}=\Zzt(-z)$,  $\tilde\A_z(z,\zb)=-\overline{\A_\zb(-\zb,-z)}=\tilde\A_z(-\zb,-z)$, thus obtaining fields defined smoothly in the whole region $\{z: -\half\leq \re z\leq \half\}$.

Similarly we can define another gauge in which $A_z=A_\zb=0$ and $\Zz(z) ,\Zzt(\zb)$ are analytic functions real on $\re z=\half$ by making a gauge transformation which maps $U_\half\mapsto 1$. In this gauge we may extend the fields over the whole strip $\{z: 0\leq \re z\leq 1\}$ by reflection about $\re z=\half$. If $\Zz', \Zzt'$ denote the fields in this second gauge, $\Zz'(z) = \overline{\Zz'(1-\zb)}$. If $\gam, \gamt$ are the gauge transformation that relate the two gauges we have constructed, $\Zz'(z+1)=\overline{\Zz'(-\zb)}=\overline{\gam(-\zb)\Zz(-\zb)}=
\overline{\gam(-\zb)}\Zz(z)$, for $-\half < \re z <0$. So, if we assume we can extend the definition of the gauge transformation $\gam(z)$ from $0<\re z<\half$ to $0<\re z<1$, we have, 
for $-\half < \re z<0$,
$$\Zz (z+1)=\gg_1(z)\Zz(z), \qquad\hbox{where}\quad \gg_1(z)=\gam(z+1)^{-1}\overline{\gam(-\zb)}.$$

Thus we have constructed from the solution defined on the cylinder, $C^2$, one defined on the torus, $T^2$, for which $\tau$ is pure imaginary, defined in a gauge in which $ \Zz(z)=\overline{\Zz(-\zb)}$. For $\tau$ pure imaginary, the complex conjugate of \nthorder, evaluated with $z$ replaced by $-\zb$, is
\eqn\nthorderc
{\theta\left[{{1\over n}(\bar\epsilon+2p)\atop -\bar\epsilon'}\right](nz,n\tau), \quad p=0, 1, \ldots n-1.}
This is in the space of functions \nthordexp\ if and only if $\epsilon$ is real and $\epsilon'=0$ or $1$. The general solution for  $\Zz$ for the cylinder is thus given by \nthordexp\ with these restrictions on $\epsilon, \epsilon'$.

\vfill\eject
\newsec{Quantization and Vertices}

{\sl Canonical Quantization.} The quantum theory involves the twistor fields $Y^I, Z^I, 1\leq I\leq 8$, the current algebra, $J^A$, ghost fields, $b,c$, associated with the conformal invariance, and ghosts $u,v,$  associated with the gauge invariance, and the world sheet gauge fields $\Aa_\mu, \Aat_\mu$ which have
no kinetic term. The conformal spins and contributions of the various fields to the central charge of the Virasoro algebra are shown the following table:

\vs
\hfil\hbox{
\vbox{\offinterlineskip
\halign{\vrule#&
   \strut\quad\hfil#\hfil\quad&\vrule#&
   \strut\quad\hfil$#$\hfil\quad&\vrule#&
   \strut\quad\hfil$#$\hfil\quad&\vrule#&
   \strut\quad\hfil$#$\hfil\quad&\vrule#&
   \strut\quad\hfil$#$\hfil\quad&\vrule#&
   \strut\quad\hfil$#$\hfil\quad&\vrule#&
   \strut\quad\hfil$#$\hfil\quad&\vrule#&
   \strut\quad\hfil$#$\hfil\quad&\vrule#\cr
\noalign{\hrule}
height4pt&\omit&&\omit&&\omit&&\omit&&\omit&&\omit&&\omit&&\omit&\cr
&\omit&&Y&&Z&&J^A&&u&&v&&b&&c&\cr
height4pt&\omit&&\omit&&\omit&&\omit&&\omit&&\omit&&\omit&&\omit&\cr
\noalign{\hrule}
height4pt&\omit&&\omit&&\omit&&\omit&&\omit&&\omit&&\omit&&\omit&\cr
&$U(1)$ charge&&-1&&1&&0&&0&&0&&0&&0&\cr
height4pt&\omit&&\omit&&\omit&&\omit&&\omit&&\omit&&\omit&&\omit&\cr
\noalign{\hrule}
height4pt&\omit&&\omit&&\omit&&\omit&&\omit&&\omit&&\omit&&\omit&\cr
&conformal spin, $\J$&&1&&0&&1&&1&&0&&2&&-1&\cr
height4pt&\omit&&\omit&&\omit&&\omit&&\omit&&\omit&&\omit&&\omit&\cr
\noalign{\hrule}
height4pt&\omit&&\omit&\omit&\omit&&\omit&&\omit&\omit&\omit&&\omit&\omit&
\omit&\cr
&central charge, $c$&&\multispan3{\hfil 0\hfil}&&28&&\multispan3{\hfil -2\hfil}&&
\multispan3{\hfil -26\hfil}&\cr
height4pt&\omit&&\omit&\omit&\omit&&\omit&&\omit&\omit&\omit&&\omit&\omit&
\omit&\cr
\noalign{\hrule}
}}}

The fields $Z^I, 1\leq I\leq 8$, comprise four boson fields, $\lambda^a,\mu^a, 1\leq a\leq 2$, and 
four fermion fields $\psi^M, 1\leq M\leq 4$; the gauge invariance insures that the $Z^I$ 
are effectively projective coordinates in the target space $\Cop\Pop^{3|4}$.
As we saw in section 2, in a gauge in which $A_z=\Aat_z=0$ and $ A_\zb=\Aa_\zb=0$,
$ \lambda^a(z),\mu^a(z),\psi^M(z)$ are holomorphic, and the only further effect of the gauge fields is through their topology. Since the contribution to the Virasoro central charge $c$ from the fermionic and
bosonic twistor fields cancels to zero, and the ghost fields have $c= -26 -2$,
the current algebra $J^A$ is required to have $c=28$.

The mode expansion of the basic fields takes the form 
\eqn\expand{
\Phi(z)=\sum \Phi_nz^{-n-\J}}
where $\Phi$ stands for $Y,Z, u, v, b, c,$ or $J^A$, and $\J$ denotes the conformal spin of the relevant fields. The vacuum state $\vac$ satisfies $\Phi_n\vac=0$ for $n>-\J$.The canonical commutation relations for the basic fields are 
\eqn\cancom{
[\![Z^i_m,Y^j_n]\!]=\delta^{ij}\delta_{m,-n},\quad
\{c_m,b_n\}=\delta_{m,-n},\quad\{v_m,u_n\}=\delta_{m,-n},}
where the brackets $[\![,]\!]$ denote commutators when either $i$ or $j$ is not greater than 4 and anticommutators when both $i$ and $j$ are greater than 4; and 
\eqn\curralg{
[J^A_m,J^B_n]=i{f^{AB}}_CJ^C_{m+n}+km\delta_{m,-n}\delta^{AB}.}
These lead to the normal ordering relations
\eqn\normord{
Z^i(z)Y^j(\zeta)=\,:Z^i(z)Y^j(\zeta):+{\delta^{ij}\over z-\zeta},}
and similar relations for $c, b$ and $v, u$.

The Virasoro algebra is given by
\eqn\viralg{
L(z)=-\sum_j:Y^j(z)\partial Z^j(z):-:u(z)\partial v(z):+2:\partial c(z)b(z):-:\partial b(z)c(z):+L^J(z),}
where the moments of $L^J(z)$ constitute the Virasoro algebra associated with the current algebra.

The BRST current is 
\eqn\qbrst{
Q(z) = c(z)\tilde L(z) + v(z)J(z) -:c(z)b(z)\partial c(z):
+{\scriptstyle {3\over 2}}\partial^2c(z),}
where 
$$\tilde L(z) = -\sum_j :Y^j(z)\partial Z^j(z): -:u(z)\partial v(z): + L^J(z).$$
$Q(z)$ is a primary conformal field of dimension one with respect to
$L(z)$, and the BRST charge $Q_0$ satifies $Q_0^2 = 0$,
where $Q(z) = \sum_n Q_n z^{-n-1}$.
$ L(z)$ generates a Virasoro algebra with total central charge zero.

{\sl Gauge Invariance.} The current associated with the gauge transformation is 
\eqn\gcurrent{
J(z) = -P(z)=-\sum_{j=1}^8 P^j(z),}
where
\eqn\pyz{
P^j(z)=:Y^j(z)Z^j(z):\,=\sum_ma_m^jz^{-m-1}\quad\hbox{for each } j;}
then
\eqn\aYZcr{
[a_m^i,a_n^j]=\epsilon^i\delta^{ij}m\delta_{m,-n},\qquad
[a_m^i,Y^j_n]=Y_{m+n}^j\delta^{ij},\qquad [a_m^i,Z^j_n]=-Z_{m+n}^j\delta^{ij},}
where
$\epsilon^i=-1,\hbox{ if } i\leq 4,$ and $\epsilon^i=1,$ otherwise. Hence
$$
[a_m^i,Y^j(z)]=z^mY^j(z)\delta^{ij},\qquad [a_m^i,Z^j(z)]=-z^mZ^j(z)\delta^{ij}.
$$
If we introduce $e^{q_0^i}$ so that
$$
e^{q_0^i}a_0^je^{-q_0^i}=a_0^j-\epsilon^i\delta^{ij},
$$
where $\epsilon^i$ is as in \aYZcr, the normal ordered products
\eqn\nop{
\nox e^{\pm X^j(z)}\nox\equiv e^{\pm q^j_0}\exp\left\{\mp\sum_{n< 0}{1\over n}a_n^jz^{-n}\right\}\exp\left\{\mp\sum_{n> 0}{1\over n}a_n^jz^{-n}\right\}z^{\pm a_0^j},}
define fermion fields, which can be identified with $Y^j(z)$, $Z^j(z)$ for $j\geq 5$. 
In general, formally,
$$
X^j(z)=q_0^j+a^j_0\log z-\sum_{n\ne 0}{1\over n}a_n^jz^{-n}.
$$
Although,  in the fermionic cases $j\geq 5$, we have an equivalence between the spaces generated by $Y^j_n, Z^j_n$, on the one hand, and $e^{\pm q_0^j}, a^j_n$, on the other, 
\eqn\YZfermi{
Y^j(z)=\nox e^{X^j(z)}\nox,\qquad Z^j(z)=\nox e^{-X^j(z)}\nox,\qquad j\geq 5,}
in the bosonic cases $j\leq 4$,
we need to supplement $e^{\pm q_0^j}, a^j_n$ by two fermionic fields $\xi^j(z), \eta^j(z)$,
$$
 \{\xi_m^i,\eta_n^j\}=\delta^{ij}\delta_{m,n};\qquad \eta_n\vac=0,\,n\geq0,
\quad \xi_n\vac=0,\,n\geq1.
$$
Note that in this case $\xi,\eta, e^{\pm q_0}, a$ generate a larger
space than the fields $Y,Z$.
Then we can write
\eqn\YZbose{
Y^j(z)=\nox e^{-X^j(z)}\nox\partial\xi^j(z),\qquad Z^j(z)=\nox e^{X^j(z)}\nox
\eta^j(z),\qquad j\leq 4.}

If $g(z)$ is an analytic function, representing a gauge transformation, defined in some annular neighborhood 
of the unit circle, $|z|=1$, with winding number $d$ as $z$ goes round the 
unit circle, 
\eqn\galphaa{
g(z)=z^de^{-\alpha(z)}, \qquad \alpha(z)= \sum_{n=-\infty}^\infty 
\alpha_nz^{-n} =\alpha^<(z)+\alpha_0+\alpha^>(z),}
where
\eqn\galphab{
\alpha^<(z)= \sum_{n>0}^\infty \alpha_nz^{-n},\qquad \alpha^>(z)= 
\sum_{n<0}^\infty \alpha_nz^{-n}.}
Defining $P(z)$ as in \gcurrent, with 
$$
P(z)=\sum_n a_nz^{-n-1}, \qquad e^{\pm q_0}=\prod_{j=1}^8e^{\pm q_0^j},
$$
let
$$
P[\alpha]={1\over 2\pi i}\oint P(z)\alpha(z) dz= \sum_{-\infty}^\infty a_n
\alpha_{-n}
=P[\alpha^<]+ a_0 \alpha_0+P[\alpha^>],
$$
$$
P[\alpha^<]= \sum_{n<0}\alpha_{-n}a_n,\qquad P[\alpha^>]=\sum_{n>0}
\alpha_{-n}a_n.
$$
So, if
\eqn\Ug{
U_g= e^{dq_0}e^{P[\alpha]}=e^{dq_0}e^{P[\alpha^<]}e^{a_0
\alpha_0}e^{P[\alpha^>]},}
noting that the $a_n$ commute among themselves,
$$
U_gY^j(z)U_g^{-1}=g(z)^{-1}Y^j(z),
\qquad U_gZ^j(z)U_g^{-1}=g(z)Z^j(z).
$$

The vertex operators $V_j(z_j)$ are gauge invariant if $[a_n,V_j(z_j)]=0$, so that
\eqn\Vgtrans{
e^{P[\alpha]}V_j(z_j)e^{-P[\alpha]}=V_j(z_j),}
we have
\eqn\vevinv{
\bvac U_gV_1(z_1)V_2(z_2)\ldots V_n(z_n)\vac =
\bvac e^{dq_0}V_1(z_1)V_2(z_2)\ldots V_n(z_n)\vac,}
showing that the winding number, the topology of the gauge transformation, is the 
only part affecting the amplitude. The winding number $d$ can be regarded as labeling different 
instanton sectors, which one needs to sum over, described by the operator $e^{dq_0}$.

Because the trace
$$\eqalign{
\tr\left(a_mV_1(z_1)V_2(z_2)\ldots V_n(z_n)w^{L_0}\right)
&=w^m\tr\left(V_1(z_1)V_2(z_2)\ldots V_n(z_n)w^{L_0}a_m\right)\cr
&=w^m
\tr\left(a_mV_1(z_1)V_2(z_2)\ldots V_n(z_n)w^{L_0}\right),\cr}$$
we have
 $$
\tr\left(a_mV_1(z_1)V_2(z_2)\ldots V_n(z_n)w^{L_0}\right)=0,\qquad\hbox{ if } 
m\ne0, $$ so that
\eqn\trinv{
\tr\left(U_gV_1(z_1)V_2(z_2)\ldots V_n(z_n)w^{L_0}\right)
=\tr\left(e^{dq_0} u^{a_0}V_1(z_1)V_2(z_2)\ldots V_n(z_n)w^{L_0}\right),}
showing that the trace depends on the instanton number $d$ and $u^{a_0}=e^{\alpha_0a_0}$. It is necessary to sum over $d$ and average over $u=e^{\alpha_0}$ with respect to the invariant measure $du/u=d\alpha_0$. The evaluation of such traces is discussed in Appendix C.

{\sl Scalar Products.} The reality conditions on $Y^I, Z^J$ imply that 
\eqn\realZ{\left(Z^J_n\right)^\dagger=Z^J_{-n},}
\eqn\realY{(Y^J_n)^\dagger = - Y^J_{-n}\quad\hbox{for}\quad 1\le J\le 4,\qquad
(Y^J_n)^\dagger = Y^J_{-n}\quad\hbox{for}\quad5\le J\le 8.}

It follows from \YZfermi\ and \YZbose\ that
\eqn\edq{
Y^I_{n-d}e^{dq_0}=e^{dq_0}Y^I_n,\qquad Z^I_{n+d}e^{dq_0}=e^{dq_0}Z^I_n\qquad
\hbox{for } 1\leq I\leq 8.}

The modes $Y^I_n, Z_n^I$ satisfy the vacuum conditions 
\eqn\vacann{Y^I_n\vac=0, \quad n\geq 0, \qquad Z^I_n\vac=0,\quad n\geq 1.}

If we take a single {\it fermionic} component of $Y^I, Z^I$ in isolation ({\it i.e.} $5\leq I\leq 8$), denoted $Y,Z$, scalar products can be evaluated from the basic relation $\langle 0|Z_0\vac=1$. With $e^{dq_0}$ defined for this component similarly to the above for integral $d$, we can show that $Z_0\vac= e^{-q_0}\vac$, and, more generally, $Z_{1-d}\ldots Z_0\vac = e^{-dq_0}\vac$, for positive integers $d$, so that
\eqn\edqsp{\langle 0|e^{dq_0}Z_{-d}\ldots Z_0\vac=1,}
and $\langle 0|e^{dq_0}Z_{-n_1}\ldots Z_{-n_m}\vac=0$ for other products $Z_{-n_1}\ldots Z_{-n_m}$
(unless $m=d+1$ and the  $n_1,\ldots,n_{d+1}$ are a permutation of $0,\ldots, d$).

For a single {\it bosonic} component of $Y^I, Z^I$ in isolation ({\it i.e.} $1\leq I\leq 4$), again denoted $Y,Z$, scalar products can be evaluated from the basic relation
\eqn\spb{
\langle 0| f(Z_0)\vac = \int f(Z_0)dZ_0,\qquad\hbox{or, equivalently,}\qquad 
\langle 0| e^{ikZ_0}\vac = \delta(k).}
In the bosonic case,  the matrix elements of $e^{q_0}$ are specified by
\eqn\ekZ{\langle 0|e^{ik'Z_0}e^{q_0}e^{ikZ_0}\vac=\langle 0|e^{q_0}e^{ik'Z_{-1}}e^{ikZ_0}\vac=\delta(k')\delta(k)}
and, analogously to \edqsp,
\eqn\edqspb{\langle 0|e^{dq_0}\exp\left\{i\sum_{j=0}^dk_jZ_{-j}\right\}\vac=\prod_{j=0}^d\delta(k_j),}
equivalently, 
\eqn\edqfn{\langle 0|e^{dq_0}f(Z_0,\ldots,Z_{-d})\vac=\int f(Z_0,\ldots, Z_{-d})dZ_0,\ldots ,dZ_{-d}.}
[Alternatively, instead of including the factor  $e^{dq_0}$, we could calculate tree diagrams
working in a twisted vacuum $|\hat 0\rangle=e^{\half dq_0}\vac$, where $Y,Z$ have modified conformal dimensions $1+\half d$ and $-\half d$, respectively. Here we will take the former approach as a basis for constructing loop contributions.]

In the cases at hand, with all eight components of $Y^I,Z^I$ present, the expressions have an overall scaling invariance under $Z^I(z)\mapsto kZ^I(z), Y^I(z)\mapsto k^{-1}Y^I(z)$, which would cause the integral over $Z^I_0, 1\leq I\leq 4$, to diverge, unless the invariant measure on this group is divided out to leave the invariant measure on the coset space. This invariant measure can be taken to be $d\gamma_S=dZ^j/Z^j$, with the various choices of $j$ giving equivalent answers and vacuum expectation value has the form $\underline{Z}$
\eqn\svev{\langle 0| f(\underline{Z}_0)\vac = \int f(\underline{Z}_0)d^4\underline{Z}_0/d\gamma_S
= \int f(\underline{Z}_0)Z^{j_0}_0d^4\underline{Z}_0/dZ^{j_0}_0}
where $\underline{Z}_0=(Z_0^1,Z_0^2,Z_0^3,Z_0^4),$ for any choice of $j_0$.  [Note that, before the vacuum expectation values are taken on the fermionic modes, $Z^I_0, 5\leq I\leq 8$, the integrand is a homogeneous function of all the $Z^I$, but after this has been done, the residual function $f(\underline{Z}_0)$ has the homogeneity property 
$f(k\underline{Z}_0)=k^{-4}f(\underline{Z}_0)$.]

{\sl Vertices.} The target space of the string theory is the twistor superspace $\Cop\Pop^{3|4}$. We use 
\eqn\Zspace{Z=\pmatrix {\lambda\cr\mu\cr\psi^\lambda\cr\psi^\mu\cr},\quad
\psi^\lambda=\pmatrix{\psi^1\cr\psi^2\cr},\quad
\psi^\mu=\pmatrix{\psi^3\cr\psi^4\cr},}
where the $\psi^I, 1\leq I\leq 4$ are anticommuting quantities and $\lambda,\mu\in\Cop^2$, to denote homogeneous coordinates in this space, identifying $Z$ and $kZ$ for nonzero $k\in\Cop$.

The vertex operator corresponding to the physical state $|\Psi\rangle =f(Z_0)J^A_{-1}\vac$ of the string, which corresponds to a gluon state, is 
\eqn\fJvertex{V(\Psi,z)=f(Z(z))J^A(z).}
The state $|\Psi\rangle$ is gauge invariant provided that $f(Z_0)$ is an homogeneous function of $Z_0$,
$f(kZ_0)=f(Z_0)$ for nonzero $k\in\Cop$, and then $V(\Psi,z)$ satisfies \Vgtrans. $f(Z_0)$ is the wave function describing the dependence of the state $\Psi$ on the mean position of the string in twistor superspace. Leaving aside the need to take account of the homogeneity of the coordinates, 
the wave function for the string being at $Z'$ would be $\prod_I\delta(Z^I(z)-Z^{\prime I})$. Thus, allowing for this, the wave function for $Z(z)$ to be at $Z'=(\pi,\omega,\theta)$ is
\eqn\Wvert{
W(z)=\int \prod_{a=1}^2\delta(k\lambda^a(z)-\pi^a)\delta(k\mu^a(z)-\omega^a)\prod_{b=1}^4
(k\psi^b(z)-\theta^b){dk\over k},}
noting that $\delta(k\psi-\theta)=k\psi-\theta$ is the form of the delta function for anticommuting variables.
[\Wvert\ is invariant under $Z(z)\mapsto k(z)Z(z)$ and separately under $Z'\mapsto kZ'$.]

To obtain the form of the vertex used by Berkovits and others \refs{\Berkovitsone-\BeWi},
 use the first delta function, $\delta(k\lambda^1(z)-\pi^1)$ to do the integration in \Wvert, to obtain
\eqn\Wverta{W(z)=\delta\left({\lambda^2(z)\over\lambda^1(z)}-{\pi^2\over\pi^1}\right)\prod_{a=1}^2\delta\left({\mu^a(z)\over\lambda^1(z)}-{\omega^a\over\pi^1}\right)\prod_{b=1}^4\left({\psi^b(z)\over\lambda^1(z)}-{\theta^b\over\pi^1}\right).}
Then multiply by $\exp(i\omega^b\bar\pi_b)$, to Fourier transform on $\omega^b$, and by
\eqn\Aexpn{A(\theta)=A_++\theta^bA_b+{1\over 2}\theta^b\theta^cA_{bc}+{1\over 3!}
\theta^b\theta^c\theta^dA_{bcd}
+\theta^1\theta^2\theta^3\theta^4A_-,}
(so that $A_{bc}$ corresponds to the six scalar bosons, $A_b$ and $A_{bcd}$ 
to the two helicity states of the four spin $\half$ fermions, and $A_\pm$ 
to the two helicity states of the spin one boson, in the $(\pm 1, 4(\pm\half), 
6(0))$ supermultiplet of N=4 Yang Mills theory) and integrate with respect to $\omega, \theta$,
yielding
$$\eqalign{
{1\over (\pi^1)^2}\delta\left({\lambda^2\over \lambda^1}-{\pi^2\over\pi^1}
\right)&\exp\left\{i{\mu^a\bar\pi_a\pi^1\over\lambda^1}\right\}\cr
&\hskip-34truemm\times\left[A_+
+{\pi^1\over\lambda^1}\psi^bA_b+\left({\pi^1
\over\lambda^1}\right)^2{1\over 2}\psi^b\psi^cA_{bc}+\left({\pi^1
\over\lambda^1}\right)^3{1\over 3!}\psi^b\psi^c\psi^dA_{bcd}+\left({\pi^1
\over\lambda^1}\right)^4\psi^1\psi^2\psi^3\psi^4 A_-\right],\cr
}$$
where $b, c, d$ are summed over.
This is essentially the vertex used by Berkovits, except that he omits 
the $A_b, A_{bc}, A_{bcd}$ terms.

The form of the vertex that it will be convenient to use in what follows is that obtained from \Wvert\ by 
Fourier transforming on $\omega^a$ and multiplying by $A(\theta)$ and integrating with respect to $\theta^a$, \eqnn\whatvert
$$\eqalignno{
\widehat W(z)= \int &{dk\over k}\prod_{a=1}^2\delta(k\lambda^a(z)-\pi^a)
e^{ik\mu^b(z)\bar\pi_b}\cr
&\hskip-8truemm\times\left[A_+ +k\psi^bA_b+{k^2\over 2}\psi^b\psi^cA_{bc}+{k^3\over 3!}
\psi^b\psi^c\psi^dA_{bcd}
+k^4\psi^1\psi^2\psi^3\psi^4A_-\right],&\whatvert}$$
where $\psi^b\equiv\psi^b(z)$.  So, the vertices for the  negative  and  positive 
helicity gluons are
\eqn\vneg{
V^A_-(z)=  
\int  {dk}k^3\prod_{a=1}^2\delta(k\lambda^a(z)-\pi^a)
e^{ik\mu^a(z)\bar\pi_a}J^A(z)
\psi^1(z)\psi^2(z)\psi^3(z)\psi^4(z)
}
and
\eqn\vpos{
V^A_+(z)=  
\int  {dk\over k}\prod_{a=1}^2\delta(k\lambda^a(z)-\pi^a)
e^{ik\mu^a(z)\bar\pi_a}J^A(z),
}
 respectively.

{\sl Cohomology.}
The standard argument that only the states in the cohomology of $Q$ contribute 
to a suitable trace over a space $\H$, provided that  $Q^\dagger = Q$ and 
$Q^2 = 0$, can be phrased as follows. Assuming the scalar product is 
non-singular on $\H$, we can write
$\H=\Phi\oplus\N\oplus\tilde\N$, with the following holding: $\N=\im Q$; 
$\Phi\oplus\N=\ker Q$; $\N$, $\tilde\N$ are null and orthogonal to $\Phi$; 
and $\N=Q\tilde\N$. The cohomology of 
$Q=\ker Q/\im\,Q\cong \Phi$. Taking bases $|e_m\rangle$ for $\Phi$, 
$|n_i\rangle$ for $\N$ and $|\tilde n_i\rangle$ for $\tilde\N$, with 
$\langle e_m|e_n\rangle=\epsilon_m\delta_{mn}$,
 $\langle n_i|\tilde n_j\rangle
=\delta_{ij}$, $\langle n_i|n_j\rangle=\langle \tilde n_i|\tilde n_j\rangle
=\langle n_i|e_m\rangle=\langle \tilde n_i|e_m\rangle=0$,
then the resolution of unity is
$$1=\sum\epsilon_m|e_m\rangle\langle e_m|+\sum |n_i\rangle\langle 
\tilde n_i|+\sum |\tilde n_i\rangle\langle  n_i|.$$

$Q|\tilde n_i\rangle$ is a basis for $\N$, so 
$|n_i\rangle=M_{ij}Q|\tilde n_j\rangle$ 
and $\delta_{ik}=M_{ij}\langle\tilde n_k|Q|\tilde n_j\rangle$. 
Then $M^T$ is the 
inverse of the matrix $\langle\tilde n_i|Q|\tilde n_j\rangle$,
$\overline M_{ij}=M_{ji}$, and
$$1=\sum\epsilon_m|e_m\rangle\langle e_m|+\sum Q|\tilde n_j\rangle M_{ij} 
\langle 
\tilde n_i|+\sum |\tilde n_j\rangle M_{ij}\langle  \tilde n_i|Q.$$

Then if we consider a trace over $\H$,
$\hbox{Tr}_\H\left(A(-1)^F\right)$, of an operator $A$, which commutes with $Q$ and anticommutes with
 $(-1)^F$ ,
$$\eqalign{
\hbox{Tr}_\H\left(A(-1)^F\right)&=\sum\epsilon_m\langle
e_m|A(-1)^F|e_m\rangle+\sum M_{ij} \langle \tilde n_i|A(-1)^FQ|
\tilde n_j\rangle+\sum  M_{ij}\langle  \tilde n_i|QA(-1)^F|\tilde n_j\rangle\cr
&=\sum\epsilon_m\langle e_m|A|e_m\rangle=\hbox{Tr}_\Phi(A).\cr}$$
provided that $F=0$ on $\Phi$.

We are interested in $Q=Q_0$, the zero mode of the BRST current \qbrst, 
and $F$ is the fermion number for the ghost fields, so that $(-1)^F$ 
does indeed anticommute with $Q$.

If we consider 
\eqn\Aone{A=e^{dq_0} 
\int \prod_{r=1}^n d\rho_r V^{A_r}_{\epsilon_r}(\rho_r) w^{L_0}}
where the vertex operators $V^A_\epsilon(\rho)$ are given by \vpos\ and \vneg, 
$Q$ commutes
with the integral of the product vertices as consequence of their having 
conformal spin 1 and $U(1)$ charge $0$, 
but it does not commute with $e^{dq_0}$,
$$ [ Q, e^{dq_0}] = d\sum_n c_{-n} a_n e^{2q_0}.$$ 
To correct for this, replace $e^{dq_0}$ by $e^{dq_0}e^{-d\Sigma_n c_{-n}u_n}$ 
in \Aone,
\eqn\Atwo{A=e^{dq_0} e^{-d\Sigma_n c_{-n}u_n}
\int \prod_{r=1}^n d\rho_r V^{A_r}_{\epsilon_r}(\rho_r) w^{L_0},}
ensuring that only states in the cohomology of $Q$ contribute to the 
trace $\tr(A(-1)^F)$. A similar instanton number changing operator is used 
in a different context in \refs{\NB}.
The inclusion of the factor $e^{-d\Sigma_n c_{-n}u_n}$ 
does not change the 
value of this trace  as we can see by expanding it in a power series.

\vfil\eject
\newsec{Tree amplitudes}

In this section we compute the N-gluon MHV twistor string tree amplitudes,
first with oscillators, and then compare with the path integral method,
in preparation for computing the loop amplitude, which will be given
in terms of a factor times the tree amplitude.
The three-point trees were derived in \refs{\BeMo},
and various N-point trees were computed in \refs{\BeWi} using aspects of Witten's twistor string theory
\refs{\Wittenone}.

The $n$-point tree amplitude, corresponding to gluons, in instanton sector $d$ is given by
\eqn\ngtree{
\A_n^\tree =\int \bvac e^{dq_0}V^{A_1}_{\epsilon_1}(z_1)
V^{A_2}_{\epsilon_2}
(z_2)\ldots V^{A_n}_{\epsilon_n}(z_n)\vac\prod_{r=1}^ndz_r\Big/d\gamma_M d\gamma_S}
where $d\gamma_M$ is the invariant measure on the M\"obius group $d\gamma_S$ is the invariant measure on the group of scale transformations on $Z$ (as in \svev), and $V^A_\epsilon$ is given in
\vpos\ or \vneg\ as $\epsilon=\pm$.
It follows directly from these expressions and from \edqsp\ that the $n$-point tree amplitude for gluons vanishes unless the number of negative helicity gluons is $d+1$.

{\sl The $d=0$ Sector.} If  $d=0$, we may take $\epsilon_1=-$ and all  other $\epsilon_r=+$. In the vacuum expectation value $\bvac V^{A_1}_{-}(z_1)
V^{A_2}_{+}(z_2)\ldots V^{A_n}_{+}(z_n)\vac$ the $Z^I(z_r)$ can be replaced by the constant $Z^I_0$, because all the $Z^I_n$,  for $n\ne 0$, can be removed by annihilation on the vacuum on either left or right. The resulting expression 
involves integrals over the zero modes, $\lambda^1_0,\lambda^2_0,\mu^1_0,\mu^2_0$ of $Z^I$, $1\leq I\leq 4$, and the $n$ scaling variables $k_r$ associated with the $n$ vertices as in \vpos\ and \vneg.

Leaving aside the current algebra from the vacuum expectation value of the $J^{A_r}(z_r)$, and $d\gamma_M$,
the amplitude is proportional to
\eqnn\treemppn
$$\eqalignno{
\int k_1^4&\prod_{r=1}^n{dk_r\over k_r}\prod_{r=1}^n\prod_{a=1}^2
\delta({\pi_r}^a-k_r\lambda^a)e^{ik_r\mu^a\bar\pi_{ra}}
\prod_{a=1}^2d\lambda^ad\mu^a/d\gamma_S \cr
&=\int k_1^4\prod_{r=1}^n{dk_r\over k_r}\prod_{r=1}^n\prod_{a=1}^2
\delta({\pi_r}^a-k_r\lambda^a)\prod_{a=1}^2\delta\left(\sum_{r=1}^nk_r\bar\pi_{ra}\right)
\prod_{a=1}^2d\lambda^a/d\gamma_S &\treemppn\cr}$$
involves $n+1$ integrals from $d^nkd^2\lambda /d\gamma_S$ and $2n+2$ delta functions, leaving 
in effect $n+1$ delta functions after the integrals have been done. For $n>3$, this exceeds the number required for momentum conservation and the amplitude vanishes.

So for $d=0$, we need only consider $n=3$; then the amplitude is
\eqn\treempp{
\A_{-++}^\tree=\int k_1^4\prod_{r=1}^3{dk_r\over k_r}\prod_{r=1}^n\prod_{a=1}^2
\delta({\pi_r}^a-k_r\lambda^a)
f^{A_1A_2A_3}\prod_{a=1}^2\delta\left(\sum_{r=1}^nk_r\bar\pi_{ra}\right)d\lambda^a /d\gamma_S,}
where the invariant measure of the M\"obius group, $d\gamma_M$, which
replaces the three ghost fields $\langle 0 |\prod_{r=1}^3 c(z_r)|0\rangle$,
has cancelled against the contribution from the tree level current algebra
correlator that we have normalized as
$\langle 0|J^A_1(z_1)J^A_2(z_2)J^A_3(z_3)|0\rangle = f^{A_1A_2A_3}
(z_1-z_2)^{-1}(z_2-z_3)^{-1}(z_3-z_1)^{-1}$. The current algebra arises from 
the $c=28$, $S_G$ sector of the world sheet theory. The structure constants 
$f^{ABC}$ supply
the non-abelian gauge group for the $D=4, N=4$ Yang Mills theory, as
explained in \refs{\Berkovitsone}.

Now, using the three $\delta({\pi_r}^1-k_r\lambda^1)$ to perform the $k_r$ integrals:
$$\A_{-++}^\tree= \prod_{a=1}^2\delta\left(\sum_{r=1}^3{\pi_r}^1
\bar\pi_{ra}\right){({\pi_1}^1)^3\over{\pi_2}^1{\pi_3}^1}\int\prod_{r=1}^3\delta\left({\pi_r}^2
-{\pi_r}^1{\lambda^2\over\lambda^1}\right)
{d\lambda^2\over\lambda^1}f^{A_1A_2A_3},$$
where we have also made the choice $d\gamma_S=d\lambda^1/\lambda^1$ in line with \svev.
Note that, alternatively, we could have used the $\delta({\pi_r}^2-k_r\lambda^2)$ delta functions to 
perform the $k_r$ integrals obtaining 
$$
\delta\left(\sum_{r=1}^3{\pi_r}^2\bar\pi_{ra}\right)\quad\hbox{ rather than } 
\quad\delta\left(\sum_{r=1}^3{\pi_r}^1\bar\pi_{ra}\right);
$$
together these four delta functions correspond to momentum conservation. To draw this out as an explicit overall factor, note that 
$$
\sum_{r=1}^3{\pi_r}^2\bar\pi_{ra}= \sum_{r=1}^3\left({\pi_r}^2-{\pi_r}^1
{\lambda^2\over\lambda^1}\right)\bar\pi_{ra}\quad\hbox{ when } 
\sum_{r=1}^3{\pi_r}^1\bar\pi_{ra}=0
$$
and the Jacobian to change between
$$\prod_{r=2,3}\delta\left({\pi_r}^2-{\pi_r}^1{\lambda^2\over\lambda^1}\right)
\quad\hbox{ and }\quad
\prod_{a=1,2}\delta\left(\sum_{r=1}^3\left({\pi_r}^2-{\pi_r}^1{\lambda^2\over
\lambda^1}\right)\bar\pi_{ra}\right)$$
is $\bar\pi_{21}
\bar\pi_{32}-\bar\pi_{22}\bar\pi_{31}=[2,3]$.

So, writing
$$
\prod_{a,b}\delta\left(\sum_{r=1}^3{\pi_r}^b\bar\pi_{ra}\right)\equiv
\delta^4(\Sigma\pi_r\bar\pi_r),
$$
\eqnn\treemppa
$$\eqalignno{
\A_{-++}^\tree&=\delta^4(\Sigma\pi_r\bar\pi_r)[2,3]\int {({\pi_1}^1)^3\over {\pi_2}^1{\pi_3}^1}\delta
\left({\pi_1}^2-{\pi_1}^1{\lambda^2\over\lambda^1}\right)
{d\lambda^2\over\lambda^1}f^{A_1A_2A_3}\cr
&=\delta^4(\Sigma\pi_r\bar\pi_r)[2,3] {({\pi_1}^1)^2\over {\pi_2}^1{\pi_3}^1}
f^{A_1A_2A_3}\cr
&=\delta^4(\Sigma\pi_r\bar\pi_r){[2,3]^3\over [1,2][3,1]}
f^{A_1A_2A_3}.&\treemppa\cr}$$ 

Writing the general polarization vector, $\epsilon_r$, of the $r$-th gluon, as the sum of positive and negative helicity parts, $\epsilon_r=\epsilon_r^++\epsilon_r^-$. These parts can be normalized by choosing vectors $s_{r\dot a}$ and $\bar s_{ra}$ are defined such that
${\pi_r}^a\bar s_{ra} = 1$ and ${{\bar\pi}_r}^{\dot a} s_{r\dot a} = 1$, for each $r$, and setting
\eqn\epnorm{
\epsilon^+_r=A^+_r\bar s_{ra}\bar\pi_{r\dot a},\qquad\epsilon^-_r=A^-_r\pi_{ra} s_{r\dot a}.}
Multiplying the appropriate amplitudes $A^\pm_r$ onto $\A_{-++}^\tree$, we obtain
\eqnn\treemppfa
$$\eqalignno{
\hat\A_{-++}^\tree&=\delta^4(\Sigma\pi_r\bar\pi_r){[2,3]^3\over [1,2][3,1]}
f^{A_1A_2A_3}A_1^-A_2^+A_3^+\cr
&=\delta^4(\pi_r\bar\pi_r) f^{A_1A_2A_3}
\left (\epsilon_1^-\cdot\epsilon_2^+ \epsilon_3^+\cdot p_1\,
+ \,\epsilon_2^+\cdot\epsilon_3^+ \epsilon_1^-\cdot p_2\,
+ \, \epsilon_3^+\cdot\epsilon_1^- \epsilon_2^+\cdot p_3\right ),&\treemppfa\cr
}$$
as shown in Appendix B.

{\sl The $d=1$ Sector.} 
The one-instanton sector contributes to all gluon tree amplitudes with just
two negative helicities. 
For $d=1$, we have to evaluate
\eqn\twodvev{\langle 0 |e^{q_0}V^{A_1}_-(z_1) V^{A_2}_-
(z_2) V^{A_3}_+(z_3)\ldots V^{A_n}_+(z_n)|0 \rangle,}
where, by  \edq\ and \vacann, and because $Y^I$ does not occur in $V^A(z)$, 
we can replace $ Z^I(z)$ by
$Z_0^I + z Z_{-1}^I$ (as the other terms in $Z^I(z)$ will annihilate on the vacuum on either the left 
or the right). Thus, for  $d=1$, by \edqfn, the expression \twodvev\ involves integrals over the 8 bosonic variables  $\lambda_0^a,\lambda_{-1}^a, \mu_0^a,\mu_{-1}^a$, $a=1,2$, as well as evaluating the dependence on the 8 fermionic variables $\psi_0^M,\psi_{-1}^M$, $1\leq M\leq 4$, using \edqsp.
 
The vacuum expectation value of currents 
\eqn\Jvacexp{
\langle 0|J^{A_1}(z_1)J^{A_2}(z_1)\ldots J^{A_n}(z_n)\vac}
can be written as a sum of a number of contributions, one of which has the form
 \eqn\Jvacexpa{
{f^{A_1A_2\ldots A_n}\over (z_1-z_2)(z_2-z_3)
\ldots (z_n-z_1)} }
and we use this contribution in what follows \refs{\FZ}. 
 
Thus the $d=1$ tree amplitude has the form
$$\eqalign{
\A_{--+\ldots +}^\tree&=\int \prod_{r=1}^n{dk_r\over k_r} \prod_{ra}
\delta({\pi_r}^a-k_r\lambda^a(z))e^{ik_r\mu^a(z_r)\bar\pi_{ra}}\cr
&\hskip10truemm\times k_1^4k_2^4\bvac e^{q_0}\psi^1(z_1)\psi^2(z_1)
\psi^3(z_1)\psi^4(z_1)\psi^1(z_2)\psi^2(z_2)\psi^3(z_2)\psi^4(z_2)\vac\cr
&\hskip10truemm\times \prod _ad^2
\lambda^ad^2\mu^a
{f^{A_1A_2\ldots A_n}dz_1dz_2\ldots dz_n\over (z_1-z_2)(z_2-z_3)
\ldots (z_n-z_1)}\bigg/d\gamma_Md\gamma_S\cr
}$$

Using that, for a component of the fermion field taken alone we would have
$$\bvac e^{q_0}\psi^1(z_1)\psi^1(z_2)\vac=(z_2-z_1)\bvac e^{q_0}
\psi^1_{-1}\psi^1_0\vac = z_1-z_2,$$ 
and integrating over $\mu_0^a,\mu_{-1}^a$,
$$\eqalign{
\A_{--+\ldots +}^\tree\hskip-8truemm&\cr
&=\int \prod_{r=1}^n{dk_r\over k_r} \prod_{r=1}^n\prod_{a=1}^2
\delta({\pi_r}^a-k_r\lambda^a(z_r))\prod_{a=1}^2\delta\left(\sum_{r=1}^nk_r\bar\pi_{ra}
\right)\delta\left(\sum_{r=1}^nk_rz_r\bar\pi_{ra}\right)\prod _{a=1}^2d^2\lambda^a\cr&
\hskip10truemm\times k_1^4k_2^4(z_1-z_2)^4{f^{A_1A_2\ldots A_n}dz_1dz_2
\ldots dz_n\over (z_1-z_2)(z_2-z_3)\ldots (z_n-z_1)}\bigg/d\gamma_Md\gamma_S\cr
&=\int \prod_{r=1}^n{1\over{\pi_r}^1} \prod_{r=1}^n\delta\left({\pi_r}^2
-{\lambda^2(z_r)\over\lambda^1(z_r)}{\pi_r}^1\right)\prod_{a=1}^2\delta\left(\sum_{r=1}^n
{{\pi_r}^1\bar\pi_{ra}\over\lambda^1(z_r)}\right)\delta\left(\sum_{r=1}^n
{z_r{\pi_r}^1\bar\pi_{ra}\over\lambda^1(z_r)}\right)\prod _{a=1}^2d^2\lambda^a\cr&
\hskip10truemm\times \left({{\pi_1}^1{\pi_2}^1(z_1-z_2)\over\lambda^1(z_1)
\lambda^1(z_2)}\right)^4{f^{A_1A_2\ldots A_n}dz_1dz_2\ldots dz_n\over (z_1-z_2)
(z_2-z_3)\ldots (z_n-z_1)}\bigg/d\gamma_Md\gamma_S\cr}$$
using the delta functions $\delta({\pi_r}^1-k_r\lambda^1(z_r))$ to 
perform the $k_r$ integrations again. Noting that when ${\pi_r}^2-(
\lambda^2(z_r)/\lambda^1(z_r)){\pi_r}^1=0$, for $b=1,2$
$$\sum_{r=1}^n{\pi_r}^b\bar\pi_{ra}=\sum_{r=1}^n{\lambda^b(z_r){\pi_r}^1\bar\pi_{ra}\over
\lambda^1(z_r)}=\lambda^b_0\sum_{r=1}^n{{\pi_r}^1\bar\pi_{ra}\over\lambda^1(z_r)}
+\lambda_{-1}^b\sum_{r=1}^n{z_r{\pi_r}^1\bar\pi_{ra}\over\lambda^1(z_r)},$$
the delta functions
$$\prod_{a=1}^2\delta\left(\sum_{r=1}^n{{\pi_r}^1\bar\pi_{ra}\over\lambda^1(z_r)}\right)
\delta\left(\sum_{r=1}^n{z_r{\pi_r}^1\bar\pi_{ra}\over\lambda^1(z_r)}\right)$$can 
be replaced by the momentum conserving delta function, together with a Jacobian factor,
$$(\lambda^1_0\lambda^2_{-1}-\lambda^1_{-1}\lambda^2_0)^2\delta^4(\Sigma{\pi_r}^a
\bar\pi_{rb}).$$
Then, with
\eqn\zetadef{\zeta_r={\lambda^2(z_r)\over\lambda^1(z_r)}
= {\lambda_0^2 +\lambda_{-1}^2 z \over
\lambda_0^1 + \lambda_{-1}^1z },}
using M\"obius invariance, we have that
\eqnn\treemmpp
$$\eqalignno{\A_{--+\ldots +}^\tree&=\int \prod_{r=1}^n{1\over{\pi_r}^1} 
\prod_{r=1}^n\delta({\pi_r}^2-\zeta_r{\pi_r}^1)\delta^4(\Sigma{\pi_r}^a\bar\pi_{rb})
{d^2\lambda^1d^2\lambda^2\over(\lambda^1_0\lambda^2_{-1}-\lambda^1_{-1}
\lambda^2_0)^2}\cr&\hskip10truemm\times\left[{\pi_1}^1{\pi_2}^1
(\zeta_1-\zeta_2)\right]^4{f^{A_1A_2\ldots A_n}d\zeta_1d\zeta_2\ldots 
d\zeta_n\over (\zeta_1-\zeta_2)(\zeta_2-\zeta_3)\ldots (\zeta_n-\zeta_1)}
\bigg/d\gamma_Md\gamma_S&\treemmpp\cr}$$
Noting that we can write
$${d^2\lambda^1d^2\lambda^2\over(\lambda^1_0\lambda^2_{-1}-\lambda^1_{-1}
\lambda^2_0)^2}=d\gamma_Md\gamma_S,
$$
because the left hand side is the invariant measure on the product of the M\"obius and scaling groups, and using the $\delta({\pi_r}^2-\zeta_r{\pi_r}^1)$ to do the $\zeta_r$ 
integrations,
\eqnn\treemmppf
$$\eqalignno{\A_{--+\ldots +}^\tree&=\delta^4({\pi_r}^a\bar\pi_{rb})
\left({\pi_1}^2{\pi_2}^1-{\pi_2}^2{\pi_1}^1\right)^4f^{A_1A_2\ldots A_n}
\prod_{r=1}^n\left({\pi_r}^2{\pi_{r+1}}^1-{\pi_{r+1}}^2{\pi_r}^1\right)^{-1}\cr
&=\delta^4({\pi_r}^a\bar\pi_{rb})
{\langle 1,2\rangle^4f^{A_1A_2\ldots A_n}\over\langle 1,2\rangle\langle 2,3
\rangle\ldots\langle n,1\rangle},&\treemmppf\cr}$$
with $\pi^a_{n+1}\equiv \pi^a_1$.

Of  course, this result could also be obtained within the path integral framework.
From this point of view, after integrating out the $Y^{I\mu}$ fields, the $d=1$ sector
path integral is 
\eqnn\pathA
$$\eqalignno{A^{\rm tree}&=
\sum_{d=1} \int D Z_I \,\,\, \delta
((\partial_{\bar z} - i A_{\bar z} ) Z^I)\cr
&\hskip35pt\cdot \int \prod_{i=1}^n dz_i
\int D\phi_G e^{-S_G} J^{A_1}(z_1) J^{A_2}(z_2)\ldots J^{A_n}(z_n)\cr
&\hskip40pt\cdot
\prod_{r=1}^n\,\,{dk_r\over k_r}\prod_{r,a}
\delta({\pi_r^a - k_r\lambda^a(z_r)})
\,\,e^{ik_r\mu^{a}(z_r)\bar\pi_{ra}}\,\cr
&\hskip43pt \cdot\left [A_{+r} + k_r^4
\psi^1(z_r)\psi^2(z_r)\psi^3(z_r)\psi^4(z_r)
A_{-r}\right]\bigg/d\gamma_M d\gamma_S.&\pathA\cr}$$
As discussed in section 2, we work in a gauge where the gauge
potentials are zero, so the path satisfies $\partial_{\bar z} Z^I = 0$,
and $Z^I(z) =  Z_0^I + Z_1^I z$ from 
\dgs\
.
Computing the path integral by replacing 
$D Z_I \,\,\, \delta
((\partial_{\bar z} - i A^{(d=1)}_{\bar z} ) Z^I)$ with 
$ \prod_{I=1}^8 dZ_0^I dZ_1^I$, and
performing the integrals over $Z_0^I, Z_1^I$,
where the fields $\lambda^a(z),\mu^a(z),\psi^M(z)$
are now given by the solutions $Z^I(z) =Z_0^I + z Z_1^I$, we see that
evaluating the path integral gives
the result obtained from the canonical quantization.
Identifying the integration
variables $Z_0^I,Z_1^I$
with the variables $\lambda^a_{0,-1},\mu^a_{0,-1},\psi^M_{0,-1}$, and
performing the Grassmann integration $\int d\psi_0 \psi_0 = 1,
\int d\psi_{-1} \psi_{-1} = 1$, for each $M$, we find that
for the two negative helicities in the positions $1,2$, we obtain \treemmppf,
together with a factor of $A_{-1}A_{-2}A_{+3}\ldots A_{+n}$ corresponding to the polarizations 
in \pathA.

\vfill\eject

\newsec{Loop Amplitudes}

The $n$-point loop amplitude is a sum over contributions corresponding to instanton numbers $d$.
We write the integrand of such a contribution as a product of factors:
\eqn\loopnd{
\A^\floop_{n,d}=\int\A^{\lambda\mu}_{n,d}\A^{\psi}_{n,d}\A^{J^A}_n 
\A^{\hbox{\small ghost}}{d\alpha_0\, d\tau\over 2\pi\im\tau}\prod_{r=1}^n\rho_rd\nu_r,
\qquad\rho_r=e^{2\pi i\nu_r},\quad w=e^{2\pi i \tau},}
where $\A^{\lambda\mu}_{n,d}$ is the part of the integrand associated with 
the bosonic twistor fields $\lambda, \mu$; $\A^{\psi}_{n,d}$ is the part 
associated with the fermionic twistor fields $\psi$; $\A^{J^A}_n$ is the 
part associated with the current algebra $J^A$;
$\A^{\hbox{\small ghost}}$ is the part associated with the ghost fields; and $\tau$ is pure imaginary.  The integration
with respect to $\alpha_0$ is the averaging over the gauge transformation $u^{a_0}$ referred to
in \trinv; the normalization factor $2\pi\im\tau$ will be explained below.
We shall restrict our attention to MHV amplitudes. For these, the fermionic part  
${\cal A}_{n,d}^\psi$ will vanish unless $d=2$ 
and so we shall use this value below.

{\sl Twistor bosonic contribution to the loop integrand.} Using the vertices \vneg\ and \vpos, the part of the integrand for the $n$-particle loop amplitude  associated 
with the bosonic twistor fields, $\lambda, \mu$, is
\eqnn\Almnd
$$\eqalignno{
\A^{\lambda\mu}_{n,2}
&=\int \tr\left(e^{2q_0}u^{a_0}\prod_{r=1}^n
\exp\left\{ik_r\lambda^a(\rho_r)\bar\omega_{ra}
+ik_r\mu^a(\rho_r)\bar\pi_{ra}\right\}w^{L_0}\right)\cr
&\hskip30truemm\times\prod_{r=1}^n{dk_r\over k_r}\prod_{a=1}^2 e^{-i\bar\omega_{ra}{\pi_r}^a} 
d\bar\omega_{ra}/d\gamma_S,&\Almnd\cr}
$$

Here we have rewritten the delta functions $\delta(k_r\lambda^a(\rho_r)-{\pi_r}^a)$ as Fourier transforms on $\bar\omega_{ra}$ in order that we can perform the trace on the bosonic variables 
$\lambda^a$, using the relation (C.1) from Appendix C
\eqn\tracel{
\tr\left(e^{dq_0}u^{a_0}\prod_{j=1}^ne^{i\omega_jZ(\rho_j)}w^{L_0}\right)
=u^{(d+1)/2}\prod_{i=1}^d\delta\left(\sum_{j=1}^nF^d_i(\hat\rho_j,w)\omega_j\right),}
where  $\hat\rho_j=u^{-\half}\rho_j=e^{2\pi i\hat\nu_j}$, 
$\hat\nu_j=\nu_j + i\alpha_0/4\pi$,
and
\eqn\Fddef{\eqalign{
F^d_k(\rho,w)&=\sum_{n=-\infty}^\infty w^{\half (n-1)(d(n- 2)
+2 k)}
\left({\rho\over w}\right)^{d(1-n)-k}\cr
&= \rho^{d/2} w^{{d/8} -{k/4}}\,\,
\theta\left[{2k/d- 1\atop 0}\right](-d\nu,d\tau)\,,\cr}}
using the notation of \thetach, so that
\eqn\Ftwo{
F^2_1(\rho,w)=\rho\theta_3(2\nu,2\tau),\qquad F^2_2(\rho,w)=w^{-{1\over 4}}\rho\theta_2(2\nu,2\tau),}
\eqnn\Antwoa
From this we have
$$\eqalignno{\A^{\lambda\mu}_{n,2}
&=u^6\int\prod_{i,a=1}^2\delta
\left(\sum_{r=1}^nk_rF^2_i(\hat\rho_r,w)\bar\omega_{ra}\right)
\delta\left(\sum_{r=1}^nk_rF^2_i(\hat\rho_r,w)\bar\pi_{ra}\right)\cr
&\hskip30truemm\times\prod_{r=1}^n {dk_r\over k_r}\prod_{a=1}^2 
e^{-i\bar\omega_{ra}{\pi_r}^a} 
 d\bar\omega_{ra}/d\gamma_S.&\Antwoa\cr}$$
 for, putting $d=2$ in \tracel, we see that we get a factor of $u^{3\over 2}$ for each of the four bosonic components of $Z$.
 Expressing the second delta functions as Fourier transforms  on $\tilde\lambda^a_i$,
 $$\eqalignno{\hskip5truemm\A^{\lambda\mu}_{n,2}
&=u^6\int \prod_{i,a=1}^2\delta\left(\sum_{r}k_rF^2_i(\hat\rho_r,w)\bar
\pi_{ra}\right)\cr
&\times\exp\left(i\sum_{r=1}^n\sum_{a=1}^2\left[\sum_{i=1}^2k_r
\tilde\lambda^a_iF^2_i(\hat\rho_r,w)\bar\omega_{ra}-i\bar\omega_{ra}
{\pi_r}^a\right]\right)\prod_{a=1}^2d^2\tilde\lambda^a\prod_{r=1}^n {dk_r\over k_r}\prod_{a=1}^2 
 d\bar\omega_{ra}/d\gamma_S\cr
 &\hskip-10truemm=u^6\int\prod_{r=1}^n
\prod_{a=1}^2\delta\left(k_r\tilde\lambda^a(\hat\rho_r,w)-{\pi_r}^a\right)
\prod_{i,a=1}^2 \delta\left(\sum_{r=1}^nk_rF^2_i(\hat\rho_r,w)\bar\pi_{ra}
\right)
\prod_{a=1}^2d^2\tilde\lambda^a\prod_{r=1}^n{dk_r\over k_r}/d\gamma_S,\cr}$$
where we have performed the integrals over $\bar\omega_{ra}$ and set
\eqn\dlambda{\tilde\lambda^a(\hat\rho_r,w)= \tilde\lambda^a_1F^2_1(\hat\rho_r,w)+\tilde
\lambda^a_2F^2_2(\hat\rho_r,w).}
Performing the $k_r$ integrations using the $\delta\left(k_r\tilde\lambda^1
(\hat\rho_r,w)-{\pi_r}^1\right)$ delta functions,
$$
\A^{\lambda\mu}_{n,2}=u^6\int \prod_{r=1}^n{1\over{\pi_r}^1}
\delta\left({\tilde\lambda^2(\hat\rho_r,w)\over \tilde\lambda^1(\hat\rho_r,w)}
{\pi_r}^1-{\pi_r}^2\right)
\prod_{i,a=1}^2\delta\left(\sum_{r=1}^n {F^2_i(\hat\rho_r,w)\over 
\tilde\lambda^1(\hat\rho_r,w)}{\pi_r}^1\bar\pi_{ra}\right)
\prod_{a=1}^2d^2\tilde\lambda^a/d\gamma_S.
$$
As before, given the constraints $\tilde\lambda^2(\hat\rho_r,w){\pi_r}^1=\tilde
\lambda^1(\hat\rho_r,w){\pi_r}^2$, we can replace
$$
\prod_{i,a=1}^2\delta\left(\sum_{r=1}^n {F^2_i(\hat\rho_r,w)\over \tilde
\lambda^1(\hat\rho_r,w)}{\pi_r}^1\bar\pi_{ra}\right)\quad\hbox{ by }\quad 
(\tilde\lambda^1_1\tilde\lambda^2_2-\tilde\lambda^1_2\tilde\lambda^2_1)^2
\delta^4 (\Sigma {\pi_r}\bar\pi_{r}),$$
so that
\eqn\Almnda{
\A^{\lambda\mu}_{n,2}=\delta^4 (\Sigma {\pi_r}\bar\pi_{r})u^6\int 
(\tilde\lambda^1_1\tilde\lambda^2_2-\tilde\lambda^1_2\tilde\lambda^2_1)^2
\prod_{r=1}^n{1\over{\pi_r}^1}\delta\left(\tx(\hat\nu_r,\tau)
{\pi_r}^1-{\pi_r}^2\right)
\prod_{a=1}^2d^2\tilde\lambda^a/d\gamma_S,}
where
\eqnn\txdef
$$\eqalignno{\tx(\hat\nu,\tau)={\tilde\lambda^2(\hat\rho,w)\over\tilde\lambda^1
(\hat\rho,w)}&={\tilde\lambda^2_1F^2_1(\hat\rho,w)+\tilde\lambda^2_2F^2_2(\hat\rho,w)
\over\tilde\lambda^1_1F^2_1(\hat\rho,w)+\tilde\lambda^1_2F^2_2(\hat\rho,w)}\cr
&={\tilde\lambda^2_1\xi(\hat\nu,\tau)+\tilde\lambda^2_2\over\tilde\lambda^1_1\xi
(\hat\nu,\tau)+\tilde\lambda^1_2},
\qquad \hbox{for }\xi(\hat\nu,\tau)={F^2_1(\hat\rho,w)\over F^2_2(\hat\rho,w)}.&\txdef
\cr}$$
Now we use the delta functions $\delta\left(\tx(\hat\nu_r,\tau){\pi_r}^1-{\pi_r}^2
\right)$ to do the integrations over $\nu_r$ in \loopnd:
\eqnn\loopnda
$$\eqalignno{
\A^\floop_{n,2}=\delta^4(\Sigma
{\pi_r}&\bar\pi_{r})\int  u^6
\prod_{r=1}^n{1\over\left({\pi_r}^1\right)^2\tx'(\hat\nu_r,\tau)}\cr
&\times(\tilde\lambda^1_1\tilde\lambda^2_2-\tilde\lambda^1_2\tilde\lambda^2_1)^2
\rhoP \A^{\psi}_{n,d}\A^{J^A}_n 
\A^{\hbox{\small ghost}}{d\alpha_0\, d\tau\over2\pi\im\tau}\prod_{a=1}^2d^2
\tilde\lambda^a/d\gamma_S,
&\loopnda\cr}$$
where $\rhoP = \prod_{r=1}^n \rho_r$ and
$$
\tx'(\hat\nu,\tau)\equiv{\partial\tx(\hat\nu,\tau)\over\partial\hat\nu}.
$$
It is to be understood that in the rest of the integrand $k_r$ and $\nu_r$ 
are determined by
\eqn\krtx{
k_r={{\pi_r}^1\over\tilde\lambda^1(\hat\rho_r,w)},\qquad \tx(\hat\nu_r,\tau)
={{\pi_r}^2\over{\pi_r}^1},}
so the second equation determines $\nu_r$ as a function of 
$ \tilde\lambda^a_i,\tau,  \alpha_0$ and $\pi_r^2/\pi_r^1$.

{\sl Twistor fermionic contribution to the loop integrand.}
Now consider the fermionic part of the integrand, 
\eqn\Apsind{
\A^{\psi}_{n,2}
=k_1^4k_2^4\tr\hskip-3pt\left(e^{2q_0}u^{a_0}(-1)^{a_0}\psi^1(\rho_1)
\psi^2(\rho_1)\psi^3(\rho_1)\psi^4(\rho_1)
\psi^1(\rho_2)\psi^2(\rho_2)\psi^3(\rho_2)\psi^4(\rho_2)w^{L_0}\right).}
If we consider one component of the fermion field $\psi^M(\rho)$ taken in 
isolation,
$$\eqalignno{
\tr\left(e^{2q_0}u^{a_0}(-1)^{a_0}\psi^1(\rho_1)\psi^1(\rho_2)w^{L_0}\right)&
=u^{-{3\over2}}F^2_1(\rho_1,w)F^2_2(\rho_2,w)-F^2_2(\rho_1,w)F^2_1(\rho_2,w)\cr
&=-u^{-{3\over2}}(\tx(\nu'_1,\tau)-\tx(\nu'_2,\tau)){\tilde\lambda^1(\rho_1,w)\tilde\lambda^1
(\rho_2,w)\over
\tilde\lambda^1_1\tilde\lambda^2_2-\tilde\lambda^1_2\tilde\lambda^2_1}\cr
&=-{u^{-{3\over2}}\langle 1,2\rangle \over k_1k_2(\tilde\lambda^1_1\tilde\lambda^2_2-\tilde
\lambda^1_2\tilde\lambda^2_1)}\cr
}$$
so that
\eqn\Apsinda{
\A^{\psi}_{n,2}={u^{-6}\langle 1,2\rangle^4 \over
(\tilde\lambda^1_1\tilde\lambda^2_2-\tilde\lambda^1_2\tilde\lambda^2_1)^4}.}
Thus, from \loopnda, we see that the factors of $u$ cancel and
\eqn\loopndb{
\A^\floop_{n,2}=\langle 1,2\rangle^4\delta^4(\Sigma
{\pi_r}\bar\pi_{r})
\int \left[\prod_{r=1}^n{1\over\left({\pi_r}^1\right)^2\tx'(\hat\nu_r,\tau)}\right]
{\A^{J^A}_n 
\A^{\hbox{\small ghost}}\over(\tilde\lambda^1_1\tilde
\lambda^2_2-\tilde\lambda^1_2\tilde\lambda^2_1)^2} 
\rhoP {d\alpha_0\,  d\tau\over 2\pi\im\tau}\prod_{a=1}^2d^2\tilde\lambda^a/d\gamma_S.}

Since
$$\tx(\nu,\tau)={\tilde\lambda^2_1\xi(\nu,\tau)+\tilde\lambda^2_2\over\tilde
\lambda^1_1\xi(\nu,\tau)+\tilde\lambda^1_2},
$$
\eqnn\xitxi
$$\eqalignno{
\prod_{r=1}^n{\xi'(\hat\nu_r,\tau)\over \xi(\hat\nu_r,\tau)-\xi(\hat\nu_{r+1},\tau)}
&=\prod_{r=1}^n{\tx'(\hat\nu_r,\tau)\over \tx(\hat\nu_r,\tau)-\tx(\hat\nu_{r+1},\tau)}\cr
&=\prod_{r=1}^n{({\pi_r}^1)^2\tx'(\hat\nu_r,\tau)\over {\pi_r}^2{\pi_{r+1}}^1 -
{\pi_{r+1}}^2{\pi_r}^1}
=\prod_{r=1}^n{({\pi_r}^1)^2\tx'(\hat\nu_r,\tau)\over \langle r,r+1\rangle}&\xitxi}
$$
Using this in \loopndb,
\eqnn\loopndc
$$\eqalignno{
\A^\floop_{n,2}&={\langle 1,2\rangle^4 
\delta^4(\Sigma{\pi_r}\bar\pi_{r})
\over \langle 1,2\rangle\langle 2,3\rangle\ldots\langle n,1\rangle}\cr
&\hskip3truemm\times\int \left[\prod_{r=1}^n{\xi(\hat\nu_r,\tau)-\xi(\hat\nu_{r+1},\tau)\over\xi'
(\hat\nu_r,\tau)}\right]{\A^{J^A}_n 
\A^{\hbox{\small ghost}}\over(\tilde\lambda^1_1\tilde
\lambda^2_2-\tilde\lambda^1_2\tilde\lambda^2_1)^2} 
\prod_{a=1}^2d^2\tilde\lambda^a{\rhoP\over d\gamma_S}{ d\alpha_0 d\tau\over2\pi\im\tau}
.&\loopndc}$$
which separates out from the rest of the amplitude the kinematic factor present in the tree amplitude.

From \Ftwo\ we have that
$$
\xi(\nu,\tau)=w^{1\over 4}{\theta_3(2\nu,2\tau)\over\theta_2(2\nu,2\tau)}.
$$
Using the relations,
$$\eqalignno{
2\theta_3(2\nu,2\tau)\theta_3(0,2\tau)&=\theta_3(\nu,\tau)^2
+\theta_4(\nu,\tau)^2,\cr
\qquad 2\theta_2(2\nu,2\tau)\theta_2(0,2\tau)&=\theta_3(\nu,\tau)^2
-\theta_4(\nu,\tau)^2,\cr}$$
we see that $\xi(\nu,\tau)$ is related by a bilinear transformation, 
whose coefficients are functions of $\tau$, to $\theta_3(\nu,\tau)^2/\theta_4
(\nu,\tau)^2$. Further, using the relation
 $$\theta_1(\nu,\tau)^2\theta_2(0,\tau)^2=
\theta_4(\nu,\tau)^2\theta_3(0,\tau)^2-\theta_3(\nu,\tau)^2\theta_4(0,\tau)^2,$$
 we see that $\xi(\nu,\tau)$ is also related by a bilinear transformation 
to the Weierstrass $\P$ function
\eqn\WP
{\P(\nu,\tau)
={\pi^2\over 3}\left[\theta_2(0,\tau)^4-\theta_4(0,\tau)^4\right]
+\left[{\theta'_1(0,\tau)\theta_3(\nu,\tau)\over\theta_3(0,\tau)\theta_1(\nu,\tau)}\right]^2.}
Using the bilinear invariance of the integrand, this implies
\eqnn\loopndd
$$\eqalignno{
\A^\floop_{n,2}&={\langle 1,2\rangle^4 
\delta^4(\Sigma{\pi_r}\bar\pi_{r})
\over \langle 1,2\rangle\langle 2,3\rangle\ldots\langle n,1\rangle}\cr
&\hskip5truemm\times\int \left[\prod_{r=1}^n{\P(\hat\nu_r,\tau)-\P(\hat\nu_{r+1},\tau)\over\P'
(\hat\nu_r,\tau)}\right]{\A^{J^A}_n 
\A^{\hbox{\small ghost}}\over(\tilde\lambda^1_1\tilde
\lambda^2_2-\tilde\lambda^1_2\tilde\lambda^2_1)^2} 
\prod_{a=1}^2d^2\tilde\lambda^a{\rhoP \over d\gamma_S}{d\alpha_0d\tau\over 2\pi\im\tau} .
&\loopndd\cr}$$
as an alternative symmetric expression. 

Regarding
$$\pmatrix{\tilde\lambda^2_1&\tilde\lambda^2_2\cr\tilde\lambda^1_1&\tilde
\lambda^1_2\cr}$$
as the matrix defining the bilinear transformation that takes 
$\xi_r=\xi(\hat\nu_r,\tau)$ to ${\pi_r}^2/{\pi_r}^1$ for $1\leq r\leq 3$, the invariant measure 
$${d^2\tilde\lambda^a\over(\tilde\lambda^1_1\tilde\lambda^2_2-\tilde
\lambda^1_2\tilde\lambda^2_1)^2}
={d\xi_1d\xi_2d\xi_3\over (\xi_1-\xi_2) (\xi_2-\xi_3) (\xi_3-\xi_1)}
d\gamma_S,
$$
\eqn\loopnde{
\A^\floop_{n,2}={\langle 1,2\rangle^4 
\delta^4(\Sigma{\pi_r}\bar\pi_{r})
\over \langle 1,2\rangle\langle 2,3\rangle\ldots\langle n,1\rangle}
\int {(\xi_3-\xi_4)\over(\xi_3-\xi_1)}
\left [ \prod_{r=4}^n {(\xi_r-\xi_{r+1})\over\xi'_r}\right]
\A^{J^A}_n \A^{\hbox{\small ghost}}\rhoP
 d\nu_1 d\nu_2 d\nu_3{d\alpha_0d\tau\over 2\pi\im\tau}.}
For  the first non vanishing amplitude,  $n=4$, noting that 
$$
{\langle 1,2\rangle\langle 3,4\rangle\over\langle 1,4\rangle\langle 3,2\rangle}
=-{s\over t}$$ 
and
$$\int{(\xi_1-\xi_2)(\xi_3-\xi_4)\over(\xi_1-\xi_4)(\xi_3-\xi_2)}\delta
\left({(\xi_1-\xi_2)(\xi_3-\xi_4)\over(\xi_1-\xi_4)(\xi_3-\xi_2)}+{s\over t}
\right)d\nu_4=
{(\xi_3-\xi_4)(\xi_4-\xi_1)\over(\xi_3-\xi_1)\xi'_4},
$$
\eqn\loopndf{
\A^\floop_{4,2}=-{\langle 1,2\rangle^4 
\delta^4(\Sigma{\pi_r}\bar\pi_{r})
\over \langle 1,2\rangle\langle 2,3\rangle\langle 3,4\rangle\langle 4,1\rangle}
{s\over t}\int \delta\left({(\xi_1-\xi_2)(\xi_3-\xi_4)\over(\xi_1-\xi_4)
(\xi_3-\xi_2)}+{s\over t}\right)\A^{J^A}_4
 \A^{\hbox{\small ghost}} \prod_{r=1}^4\rho_rd\nu_r{d\alpha_0
d\tau\over 2\pi\im\tau}.}

Similarly, we can derive a corresponding expression for the $n$-point loop,
\eqnn\loopndfg
$$\eqalignno{
\A^\floop_{n,2}&={\langle 1,2\rangle^n\delta^4(
\Sigma {\pi_r}\bar\pi_{r})
\over\langle 2,3\rangle^{n-2}\langle 3,1\rangle^{n-4}}\hskip15truemm\cr
&\hskip-5truemm\times\int \prod_{r=4}^n{1\over \langle r, 1\rangle^2}\delta\left({(\xi_r-\xi_3)
(\xi_2-\xi_1)\over(\xi_r-\xi_1)(\xi_2-\xi_3)}-{\langle r,3\rangle\langle 2,1
\rangle\over\langle r,1\rangle\langle 2,3\rangle}\right)\A^{J^A}_n 
\A^{\hbox{\small ghost}}\,
\prod_{r=1}^n\rho_rd\nu_r{d\alpha_0
d\tau\over 2\pi\im\tau}. \hskip10truemm&\loopndfg\cr}$$

Because $\hat\nu_r = \nu_r + i\alpha_0/4\pi$ and 
 the integral is invariant as $\nu_r\rightarrow\nu_r + 
\half \tau$, we see that it is periodic under $\alpha_0\rightarrow 
\alpha_0+2\pi {\rm Im}\tau$.
So to average over $\alpha_0$, we integrate over the range
$0\leq \alpha_0\leq 2\pi {\rm Im}\tau$ and divide by $2\pi {\rm Im}\tau$, which explains the normalization factor introduced in \loopnd.

{\sl Path Integral Derivation.}
In the path integral approach, 
working in the gauge $A_i = 0$,\break
we use the paths given in \nthordexp\ with $n=2$ and
$\lambda^a(\rho)=Z^a(\nu), \mu^a(\rho)=Z^{a+2}(\nu), \break
a=1,2; \psi^M(\rho)=Z^{M+4}(\nu), 1\leq M\leq 4$.

The path integral on the cylinder includes, up to normalization,
$$\eqalign{
\int\A_{n,2}^{\lambda\mu}\,\,\prod_{r=1}^n d\nu_r
&=\int \prod_{I=1}^4 dc_0^I dc_1^I \sum_{\epsilon'=0}^1\,
\int_0^2 d\epsilon \left(\prod_{r=1}^n
\exp\left\{ik_r\lambda^a(\rho_r)\bar\omega_{ra}
+ik_r\mu^a(\rho_r)\bar\pi_{ra}\right\}\right)\cr
&\hskip20pt\times
\left(\prod_{r=1}^n{dk_r\over k_r}\prod_{a=1}^2 e^{-i\bar\omega_{ra}{\pi_r}^a}
d\bar\omega_{ra}
(\prod_{r=1}^{n-1}d\nu_r)\right)\bigg/d\gamma_S.\cr}
$$
Performing the integrations over $c_0^I , c_1^I, 1\leq I\leq 4$, we
find a formula analogous to the canonical expression \Antwoa,
save that the functions $F_1^2(\rho_r, w)$ and $F_2^2(\rho_r, w)$ are 
replaced with $\theta [{{\half\epsilon}\atop \epsilon'}](2\nu_r,2\tau)$,
 and $\theta [{{{\half\epsilon} + 1}\atop \epsilon'}](2\nu_r,2\tau)\,$,
 respectively,
and the $\alpha_0$ integration is exchanged for
a sum over $\epsilon'$ and the $\epsilon$ integration.

The calculation proceeds as before to a formula similar to 
\Almnda,
except that $\tilde \xi (\nu,\tau)$ is replaced with 
$$\tilde \eta(\nu,\tau)
= {\tilde\lambda^2_1\eta(\nu,\tau)+\tilde\lambda^2_2\over\tilde\lambda^1_1\eta
(\nu,\tau)+\tilde\lambda^1_2}\quad\hbox{where}\quad
\eta(\nu,\tau)
= {\theta [{{\half\epsilon}\atop \epsilon'}]
(2\nu,2\tau)\over \theta [{{{\half\epsilon} + 1}\atop \epsilon'}]
(2\nu,2\tau)} = {\theta_3(2\nu + {\half\epsilon'} +{\half\epsilon\tau},
2\tau)\over\theta_2(2\nu + {\half\epsilon'} +{\half\epsilon\tau},
2\tau)}.$$
After integrating with respect to $\epsilon$, the resulting integrand
depends  on the differences $\nu_i-\nu_j$, and is independent of $\epsilon'$. 
Then $\tilde\eta(\nu,\tau)$ 
is essentially $\tilde\xi(\nu,\tau)$ from (5.9), since the
factor $w^{1\over 4}$ in $\xi(\nu,\tau)$ can be absorbed in 
the $\tilde\lambda^a$ integrations, and ${\epsilon\tau/2}$
can be replaced by $i\alpha_0/2\pi$,  $d\epsilon=d\alpha_0/\pi\im\tau$ 
($\alpha_0,\epsilon$ real, $\tau$ pure imaginary). 

In analogy with 
\loopnda,
we use the delta functions
$\delta\left(\tilde\eta(\nu_r,\tau){\pi_r}^1-{\pi_r}^2
\right)$ to do the integrations over $\nu_r$:
$$\eqalign{
\int\A_{n,2}^{\lambda\mu}\,\,&\prod_{r=1}^n d\nu_r\cr
&\hskip-3truemm=\delta^4(
\Sigma {\pi_r}\bar\pi_{r}) \, 2\int_0^2 d\epsilon \int
\left[\prod_{r=1}^n{1\over\left({\pi_r}^1\right)^2\tilde\eta'
(\nu_r,\tau)}\right]
(\tilde\lambda^1_1\tilde\lambda^2_2-\tilde\lambda^1_2\tilde\lambda^2_1)^2
\left[\prod_{a=1}^2d^2\tilde\lambda^a\right]\rho_\Pi {\bigg/ d\gamma_S}.\cr}
$$
In the rest of the integrand $k_r$ and $\nu_r$
are determined by
$k_r={\pi_r}^1/\tilde\lambda^1(\rho_r,w)$ and $\tilde\eta(\nu_r,\tau)
={\pi_r}^2/{\pi_r}^1.$
The one-loop integrand for the fermionic 
fields, for one permutation of helicities,  is
$$\A_{n,2}^\psi= k_1^4k_2^4 \int \prod_{M=1}^4 dc_0^{M+4} dc_1^{M+4}
\psi^M(\rho_1)\psi^M(\rho_2)={\langle 1,2\rangle^4 \over
(\tilde\lambda^1_1\tilde\lambda^2_2-\tilde\lambda^1_2\tilde\lambda^2_1)^4}.
$$
Thus we obtain \loopndc\ as before, apart from a factor of 4.

{\sl Ghost contribution to the loop integrand.} The ghost fields in the theory 
are the customary ghost fields, $b,c$,
with conformal spin $(2,-1)$, associated with the reparametrization invariance, and the ghosts for the $U(1)$ gauge fields 
$u,v$ with conformal spin $(1,0)$. 

The partition function for a general fermionic ``$b,c$'' system
with conformal dimensions $\lambda$ and $1-\lambda$ respectively is
$$\tr \big( b_0 c_0 \omega^{L_0 - {c\over 24}} (-1)^F\big)
= \omega^{-{c\over 24}} \omega^{\half \lambda ( 1-\lambda )} 
\prod_{n=1}^\infty (1-\omega^n)^2\,,$$
where the central charge is $12\lambda (1-\lambda) -2$,
and $L(z) = -\lambda\nox b(z)c'(z)\nox
+ (1-\lambda) \nox b'(z) c(z)\nox$.
So 
$$\tr \big( b_0 c_0 \omega^{L_0 - {c\over 24}} (-1)^F\big)
= \omega^{1\over 12}\prod_{n=1}^\infty (1-\omega^n)^2 = \eta(\tau)^2\,.$$
and the reparametrization and $U(1)$ ghosts each contribute this factor to the integrand:

The factor of $(-1)^F$ is included so that the ghosts have the same
periodicity as the original coordinate transformations.
See Freeman and Olive \refs{\FO}.
The $b_0,c_0$ insertion projects onto half the states
in the $b,c$ system; without this projection, the $(-1)^F$ would
force the trace to vanish. 

So the total ghost contribution is
\eqn\Aghost{{\cal A}^{\rm ghost}
= \eta(\tau)^4.}

\vfil\eject

\newsec{Current Algebra Loop}

For the final piece of the integrand of the loop amplitude, we 
compute the  one-loop amplitude,
\eqn\CAnamp{\tr(J^{a_1}(\rho_1)J^{a_2}(\rho_2)\ldots J^{a_n}(\rho_n)w^{L_0}),}
for the current algebra of an arbitrary Lie group, $G$,
\eqn\Calg{
[J^a_m,J^b_n]=if^{abc}J^c_{m+n}+km\delta^{ab}\delta_{m,-n},
\qquad J^a(\rho)=\sum_n J^a_n\rho^{-n-1}.}
We calculate this using a recursion relation, and will present
the derivation in a future publication \refs{\DG}.
Relevant discussion is also given in \refs{\Zhu,\MT}.
The amplitudes can be expressed in terms of the $\tau$-dependent invariant tensors, 
\eqn\troo{
\tr(J_0^{a_1}J_0^{a_2}\ldots J_0^{a_n}w^{L_0}),}
which themselves can be expressed as products of (constant) invariant tensors and functions of $\tau$. We note that \CAnamp\ is symmetric under simultaneous identical permutations of the $\rho_j$ and $a_j$. The structure of the loop amplitude for $n\geq 4$ is best understood by considering first the zero, two and three-point functions. (The one-point function necessarily vanishes.)  In what follows  $L_0$ denotes the zero-mode generator for 
 the Virasoro algebra associated with the current algebra.

{\sl Fermionic Representations.} We begin by considering the case where $J^a(\rho)$ is given by a fermionic representation,
\eqn\frep{
J^a(\rho)={i\over 2}T^a_{ij}b^i(\rho)b^j(\rho)\qquad
J^a_n={i\over 2}\sum_rT^a_{ij}b^i_rb^j_{n-r}}
\eqn\brep{
[T^a,T^b]=f^{abc}T^c,\qquad  \tr(T^aT^b)=-2k\delta^{ab}, \qquad b^j(\rho)=\sum_r b^j_r\rho^{-r-\half},}
where $T^a$, $1\leq a\leq \dim G$, are real antisymmetric matrices providing a real dimension $D$ representation of $G$, 
and the $b_r^i, r\in\Zop+\half,$ are Neveu-Schwarz fermionic oscillators, $\{b_r^i,b_s^j\}=\delta^{ij}\delta_{r,-s}, \quad b_r^j\vac=0,\, r>0,\quad (b_r^j)^\dagger=b_{-r}^j$.

For this representation, the zero-point function,
\eqn\chiF{\chi(\tau)=\tr\left(w^{L_0}\right)=\prod_{s=\half}^\infty(1+w^s)^D,}
while the one-point function vanishes, $\tr(J^a(\rho)w^{L_0})$, as it does for  any representation.

The $n$-point  one-loop current algebra amplitude is computed by using the usual recurrence relation, for calculating $\tr(b^{i_1}_{r_1}\ldots b^{i_n}_{r_n}w^{L_0})$ 
in the free fermion representation, obtained by moving $b^{i_n}_{r_n}$ around the trace,
$$\tr(b^{i_1}_{r_1}\ldots b^{i_n}_{r_n}w^{L_0})
={ w^{r_n}\over 1+ w^{r_n}}\sum_{m=1}^{n-1}(-1)^{m+1}\delta_{r_m,-r_n}\delta^{i_mi_n}
\tr(b^{i_1}_{r_1}\ldots b^{i_{m-1}}_{r_{m-1}}b^{i_{m+1}}_{r_{m+1}}\ldots b^{i_{n-1}}_{r_{n-1}}w^{L_0}).
$$

Using this, we obtain the two-point loop,
\eqn\ftwo{
\tr(J^a(\rho_1)J^b(\rho_2)w^{L_0})
={k\chi(\tau)\over\rho_1\rho_2}\delta^{ab}\chi_F(\nu_1-\nu_2,\tau)^2,}
where
\eqn\chiFnu{
\chi_F(\nu,\tau)=\sum_{r=\half}^\infty {e^{2\pi i r\nu}+w^re^{-2\pi i r\nu}\over 1+w^r}
={i\over 2}\theta_2(0,\tau)\theta_4(0,\tau){\theta_3(\nu,\tau)\over\theta_1(\nu,\tau)}.}
The three-point loop,
\eqn\fthree{
\tr(J^a(\rho_1)J^b(\rho_2)J^c(\rho_3)w^{L_0})
={ikf^{abc}\chi(\tau)\over \rho_1\rho_2\rho_3}\chi_F^{21}\chi_F^{32}\chi_F^{13},}
and the four-point loop,
\eqn\ffour{\eqalign{
\tr (&J^a(\rho_1)J^b(\rho_2)J^c(\rho_3)J^d(\rho_4) w^{L_0}) \cr
&= {\chi(\tau)\over\rho_1\rho_2\rho_3\rho_4}[\sigma^{abcd} \chi^{12}_F\, \chi^{23}_F\,
\chi^{34}_F\, \chi^{14}_F
 -\sigma^{abdc} \chi^{12}_F\, \chi^{24}_F\,
\chi^{34}_F\, \chi^{13}_F
-\sigma^{acbd}  \chi^{13}_F\, \chi^{23}_F\,
\chi^{24}_F\, \chi^{14}_F\cr
&\hskip15truemm+  k^2
\, \delta^{ab}\delta^{cd}\,(\chi^{12}_F)^2\, (\chi^{34}_F)^2
+ k^2
\, \delta^{ac}\delta^{bd}\,(\chi^{13}_F)^2\, (\chi^{24}_F)^2
 +  k^2
\delta^{ad} \,\delta^{bc} \,(\chi^{14}_F)^2\, (\chi^{23}_F)^2].
\cr}}
where $\chi^{ij}_F = \chi_F(\nu_j-\nu_i,\tau)$ and 
\eqn\defsigma{\sigma^{abcd}=\tr(T^aT^bT^cT^d).}

{\sl General Representations.} The amplitude \CAnamp\ can be evaluated by means of recurrence relations between functions
$$\tr\left(J_0^{a_1}J_0^{a_2}\ldots J_0^{a_m}J^{b_1}(\rho_1)J^{b_2}(\rho_2)\ldots J^{b_n}(\rho_n)w^{L_0}\right)$$
that can be established using the cyclic properties of the trace and the algebra \Calg, from which it follows that
$$[J^a_m,J^b(z)]=iz^mf^{abc}J^c(z)+kmz^{m-1}\delta^{ab}.$$

For example, in this way it follows that
\eqnn\defreca
$$\eqalignno{
\tr&\left(J^a_mJ^{a_1}(\rho_1)\ldots J^{a_n}(\rho_n)w^{L_0}\right)(1-w^m)\cr
&\hskip10truemm=i\sum_{j=1}^nf^{aa_ja_j'}\tr\left(J^{a_1}(\rho_1)\ldots J^{a_{j-1}}(\rho_{j-1})J^{a_j'}(\rho_j)J^{a_{j+1}}(\rho_{j+1})\ldots J^{a_n}(\rho_n)\right)\rho_j^m\cr
&\hskip15truemm+km\sum_{j=1}^n\delta^{aa_j}\tr\left(J^{a_1}(\rho_1)\ldots J^{a_{j-1}}(\rho_{j-1})J^{a_{j+1}}(\rho_{j+1})\ldots J^{a_n}(\rho_n)\right)\rho^{m-1}_j&\defreca\cr
}$$
and, hence,
\eqnn\defrecb
$$\eqalignno{
\tr&\left(J^a(\rho)J^{a_1}(\rho_1)\ldots J^{a_n}(\rho_n)w^{L_0}\right)\cr
&\hskip0truemm=\rho^{-1}\tr\left(J^a_0J^{a_1}(\rho_1)\ldots J^{a_n}(\rho_n)w^{L_0}\right)\cr
&\hskip0truemm+i\sum_{j=1}^n{«\Delta_1(\nu_j\nu,\tau)\over\rho}f^{aa_j}_{\hskip10pt a_j'}
\tr\left(J^{a_1}(\rho_1)
\ldots J^{a_{j-1}}(\rho_{j-1})J^{a_j'}(\rho_j)J^{a_{j+1}}(\rho_{j+1})\ldots J^{a_n}(\rho_n)w^{L_0}\right)\cr
&\hskip0truemm+k\sum_{j=1}^n{\Delta_2(\nu_j-\nu,\tau)\over \rho\rho_j}\delta^{aa_j}\tr\left(J^{a_1}(\rho_1)\ldots J^{a_{j-1}}(\rho_{j-1})J^{a_{j+1}}(\rho_{j+1})\ldots J^{a_n}(\rho_n)w^{L_0}\right)
&\defrecb\cr
}$$
where
\eqnn\defdeltaA
\eqnn\defdeltaB
$$\eqalignno{
\Delta_1(\nu,\tau)&=\sum_{m\ne 0}{e^{2\pi im\nu}\over 1-w^m}
=\sum_{m=1}^\infty {e^{2\pi im\nu}-w^me^{-2\pi im\nu}\over 1-w^m}
={i\over 2\pi}{\theta_1'(\nu)\over\theta_1(\nu)}-{1\over 2},&\defdeltaA\cr
\Delta_2(\nu,\tau)&=\sum_{m\ne 0}{me^{2\pi im\nu}\over 1-w^m}
=\sum_{m=1}^\infty m\,{e^{2\pi im\nu}+w^me^{-2\pi im\nu}\over 1-w^m}
={1\over 2\pi i}\Delta_1'(\nu,\tau).&\defdeltaB\cr
}$$
These functions relate to the Weierstrass elliptic functions by
\eqn\deltaP{
\P(\nu,\tau) = -4\pi^2\Delta_2(\nu,\tau)-2\eta(\tau), \qquad \eta(\tau)=-{1\over 6}{\theta'''_1(0,\tau)\over\theta_1'(0,\tau)},}
\eqn\deltaZ{
\zeta(\nu,\tau)=-2\pi i\Delta_1(\nu,\tau)-\pi i+2\eta(\tau)\nu,\qquad \P(\nu,\tau)=-\zeta'(\nu,\tau).}
Using \defrecb\ and similar relations we can establish general formulae for the two-, three- and four-point one-loop functions. 

We write the zero-point (partition) function as
\eqn\chidef{\tr\left(w^{L_0}\right) \equiv\chi(\tau).}

Because it is isotropic, we can write
$$
\tr\left(J^a_0J^b_0w^{L_0}\right)
=\delta^{ab}\chi^{(2)}(\tau)\qquad\hbox{where }\quad\chi^{(2)}(\tau)={1\over\dim G} \tr\left(J^a_0J^a_0w^{L_0}\right).$$
Then the general form of the two-point current algebra loop is
\eqn\twog{\tr(J^a(\rho_1)J^b(\rho_2) w^{L_0})
= {\delta^{ab} k\chi(\tau)\over\rho_1\rho_2}  \left[\chi_F(\nu_2-\nu_1,\tau)^2
+ f(\tau)\right],}
where
\eqn\fdef{f(\tau) =  {\chi^{(2)}(\tau)\over k \chi(\tau)}
+{\theta''_3(0,\tau)\over 4\pi^2\theta_3(0,\tau)}.}

Noting that we can write
$$\tr\left(J^a_0J^b_0J^c_0w^{L_0}\right)=\half\tr\left([J^a_0,J^b_0]J^c_0w^{L_0}\right)
+\half\tr\left(\{J^a_0,J^b_0\}J^c_0w^{L_0}\right)$$
we have
$$\tr\left(J^a_0J^b_0J^c_0w^{L_0}\right)=
\half i f^{abc}\chi^{(2)}(\tau)+\half d^{abc}\chi^{(3)}(\tau),
$$
where $d^{abc}$ is a totally symmetric isotropic tensor, which may vanish, 
as it does for $SU(2)$, but not $SU(3)$.
Among the simple Lie groups, only $SU(n)$ with $n\ge 3$ has a symmetric invariant tensor of order 3 \refs{\Weyl}.
Then the general form of the three-point current algebra loop is
\eqnn\threeg
$$\eqalignno {\tr(&J^a(\rho_1)J^b(\rho_2) J^c(\rho_3) w^{L_0})\cr
&={i kf^{abc} \chi(\tau) \over\rho_1\rho_2\rho_3}\left\{
\chi_F^{21}\chi_F^{32}\chi_F^{13} -{i\over 2\pi} \left(\zeta^{21}
+ \zeta^{32} + \zeta^{13}\right)
f(\tau)\right\}+ {d^{abc} \chi^{(3)}(\tau)\over2\rho_1\rho_2\rho_3},&\threeg\cr}$$
where $\zeta^{ij}=\zeta(\nu_j-\nu_i)$. The recurrence relations do not manifestly maintain the permutation symmetry of the loop amplitudes but the final result is necessarily symmetric and can be put into this form.

The general form of the four-point loop can be put into the symmetric form
\eqnn\fourg
$$\eqalignno {\tr(J^a(\rho_1)&J^b(\rho_2) J^c(\rho_3) J^d(\rho_4)w^{L_0})\rho_1\rho_2\rho_3\rho_4\cr
= &\Big\{
\delta^{ab}\delta^{cd}\, \left (k^2 \chi(\tau)\,
[(\chi_F^{12})^2+ f(\tau)] [(\chi_F^{34})^2+ f(\tau)]-\chi^{(2)}(\tau)^2/\chi(\tau)\right)\cr
&+ \delta^{ac}\delta^{bd}
\, \left( k^2 \chi(\tau)\,
[(\chi_F^{13})^2+ f(\tau)] [(\chi_F^{24})^2+ f(\tau)]-\chi^{(2)}(\tau)^2/\chi(\tau)\right)\cr
&\hskip0pt + \delta^{ad}\delta^{bc}
\, \left(k^2 \chi(\tau)\,
[(\chi_F^{14})^2+ f(\tau)] [(\chi_F^{23})^2+ f(\tau)]-\chi^{(2)}(\tau)^2/\chi(\tau)\right)\cr
& + \tr (J^a_0 J^b_0 J^c_0J^d_0 w^{L_0})_{\bf S}\cr
&- {1\over 96}\left (\sigma^{abcd} + \sigma^{adcb}+ \sigma^{acdb} + \sigma^{abdc} +
\sigma^{adbc} +  \sigma^{acbd}\right)
\chi(\tau)\theta_2^4(0,\tau)\theta_4^4(0,\tau)\cr
& - \left (\sigma^{abcd} 
+ \sigma^{adcb}
\right)\half \chi(\tau)\,\left\{
\chi_F^{12}\chi_F^{23}\chi_F^{34}\chi_F^{41} \phantom{{{\cal P}'_{24}  - {\cal P}'_{14}\over
{\cal P}_{24}  - {\cal P}_{14}}}\right.\cr
&\hskip15truemm\left.- {1\over 16\pi^2} f(\tau) 
\left({{\cal P}'_{24} - {\cal P}'_{32} \over {\cal P}_{24}  - {\cal P}_{32}}\right)
\left( {{\cal P}'_{24}  - {\cal P}'_{41}\over{\cal P}_{24}  - {\cal P}_{41}}\right )
 + {1\over 4\pi^2} f(\tau)\P_{24}\right\}\cr
& - \left(\sigma^{acdb} 
+ \sigma^{abdc} \right)
\half\chi(\tau)\left\{
\chi_F^{13}\chi_F^{34}\chi_F^{42}\chi_F^{21} \phantom{{{\cal P}'_{24}  - {\cal P}'_{14}\over
{\cal P}_{24}  - {\cal P}_{14}}}\right.\cr
&\hskip15truemm\left.- {1\over 16\pi^2} f(\tau)
\left({{\cal P}'_{24}  - {\cal P}_{32}\over{\cal P}_{24} - {\cal P}_{32}}\right)
\left({{\cal P}'_{21} - {\cal P}'_{32}\over{\cal P}_{21} - {\cal P}_{32}}\right)
+ {1\over 4\pi^2} f(\tau) {\cal P}_{32}\right\}\cr
& -  \left(  \sigma^{adbc} +  \sigma^{acbd}\right)\,\,\half \chi(\tau)\left\{
\chi_F^{14}\chi_F^{42}\chi_F^{23}\chi_F^{31}\phantom{{{\cal P}'_{24}  - {\cal P}'_{14}\over
{\cal P}_{24}  - {\cal P}_{14}}}\right.\cr
&\hskip15truemm\left. - {1\over 16\pi^2} f(\tau) \left (
{{\cal P}'_{24}  - {\cal P}'_{32}\over {\cal P}_{24}  - {\cal P}_{32}}\right )
\left( {{\cal P}'_{14} - {\cal P}'_{31}\over
{\cal P}_{14} - {\cal P}_{31}}\right)  +{1\over 4\pi^2} f(\tau) {\cal{P}}_{34}\right\}\cr
&- {i\over 4\pi}
\chi^{(3)}(\tau)\left[\sigma^{abcd} \left(\zeta^{21} + \zeta^{32} + \zeta^{43}+\zeta^{14}\right)
+ \sigma^{adcb}\left(\zeta^{41} + \zeta^{34} + \zeta^{23}+\zeta^{12}\right)\right.\cr
&\hskip21truemm+ \sigma^{acdb} \left(\zeta^{31} + \zeta^{43} + \zeta^{24}+\zeta^{12}\right)
+ \sigma^{abdc} \left(\zeta^{21} + \zeta^{42} + \zeta^{34}+\zeta^{13}\right)\cr
&\left. \hskip21truemm+ \,\sigma^{adbc}\left(\zeta^{41} + \zeta^{24} + \zeta^{32}+\zeta^{13}\right)
 +  \sigma^{acbd}
\left(\zeta^{31} + \zeta^{23} + \zeta^{42}+\zeta^{14}\right)\right]\Big\},\cr
&&\fourg\cr
}$$
where $\P_{ij}=\P(\nu_j-\nu_i,\tau)$, $\P'_{ij}=\P'(\nu_j-\nu_i,\tau)$, $\sigma^{abcd}$ is given by \defsigma\ and $\tr (J^a_0 J^b_0 J^c_0J^d_0 w^{L_0})_{\bf S}$ is the symmetrization of the trace
$\tr (J^a_0 J^b_0 J^c_0J^d_0 w^{L_0})$ over permutations of $a,b,c,d$.

We will now specialize to the case of a general representation of  $SU(2)$.
 In this case,
 \eqnn\foursut
$$\eqalignno {\tr(&J^a(\rho_1)J^b(\rho_2) J^c(\rho_3) J^d(\rho_4)w^{L_0})\rho_1\rho_2\rho_3\rho_4\cr
&=\Big\{
\delta^{ab}\delta^{cd}\, \left (k^2 \chi(\tau)\,
[(\chi_F^{12})^2+ f(\tau)] [(\chi_F^{34})^2+ f(\tau)]-\chi^{(2)}(\tau)^2/\chi(\tau)\right)\cr
&\hskip3truemm+ \delta^{ac}\delta^{bd}
\, \left( k^2 \chi(\tau)\,
[(\chi_F^{13})^2+ f(\tau)] [(\chi_F^{24})^2+ f(\tau)]-\chi^{(2)}(\tau)^2/\chi(\tau)\right)\cr
&\hskip3truemm + \delta^{ad}\delta^{bc}
\, \left(k^2 \chi(\tau)\,
[(\chi_F^{14})^2+ f(\tau)] [(\chi_F^{23})^2+ f(\tau)]-\chi^{(2)}(\tau)^2/\chi(\tau)\right)\cr
&\hskip3truemm- [\sigma^{abcd}+\sigma^{acdb}+\sigma^{adbc}]\,\, \chi(\tau)
\left({1\over 48}\theta_2^4(0,\tau)\theta_4^4(0,\tau)
- {1\over 6} {\chi^{(4)}(\tau)\over k\chi(\tau)}\right)\cr
&\hskip3truemm- \sigma^{abcd}\,\,\chi(\tau)\,\Big\{\chi_F^{12}\chi_F^{23}\chi_F^{34}\chi_F^{41} 
 +{f(\tau)\over 8\pi^2}\Big[{\cal P}_{13}+ {\cal P}_{24}\cr
&\hskip6truemm -\left(\zeta^{13}+ \zeta^{32} + \zeta^{21}\right)
\left(\zeta^{13}+ \zeta^{34} + \zeta^{41}\right)-
\left(\zeta^{24}+ \zeta^{41} + \zeta^{12}\right)
\left(\zeta^{24}+ \zeta^{43} + \zeta^{32}\right)\Big ]\Big\}\cr
&\hskip3truemm - \sigma^{acdb}\,\,\chi(\tau)\,\Big\{\chi_F^{13}\chi_F^{34}\chi_F^{42}\chi_F^{21} 
+{f(\tau)\over 8\pi^2}\Big[{\cal P}_{14}+ {\cal P}_{23}\cr
&\hskip6truemm
-\left(\zeta^{14}+ \zeta^{42} + \zeta^{21}\right)
\left(\zeta^{14}+ \zeta^{43} + \zeta^{31}\right)-
\left(\zeta^{23}+ \zeta^{31} + \zeta^{12}\right)
\left(\zeta^{23}+ \zeta^{34} + \zeta^{42}\right)\Big ]\Big\}\cr
&\hskip3truemm-  \sigma^{adbc}\,\,\chi(\tau)\,\Big\{ \chi_F^{14}\chi_F^{42}\chi_F^{23}\chi_F^{31}
 +{f(\tau)\over 8\pi^2}\Big[{\cal P}_{12}+ {\cal P}_{34}\cr
&\hskip6truemm
-\left(\zeta^{12}+ \zeta^{23} + \zeta^{31}\right)
\left(\zeta^{12}+ \zeta^{24} + \zeta^{41}\right)-
\left(\zeta^{34}+ \zeta^{41} + \zeta^{13}\right)
\left(\zeta^{34}+ \zeta^{42} + \zeta^{23}\right)\Big ]\Big \}\Big\}.\cr
&&\foursut\cr}$$

\vfill\eject

\newsec{Twistor String Loop}

We now assemble the parts for the one-loop MHV gluon amplitude of the
twistor string. The fact that the twistor string has
delta function vertices leads to the form for
the final integral that is a simple product of the loop for the twistor
fields and the current algebra loop.
We have provided several equal expressions for the twistor
field loop $\A_{n,2}^{\rm loop}$ in  \loopndc,  \loopndd\ - \loopndfg.
These forms were given for particles $1,2$ having negative helicity, and
we saw that the expressions naturally divide into two factors: a piece equal to
the kinematic factor of the tree amplitude multiplied by a function of $s$ and $t$.
For the four-gluon amplitude,
if we consider the form given in  \loopndf,  we have
$$
\A^\floop_{4,2}=-{\langle 1,2\rangle^4 
\delta^4(\Sigma{\pi_r}\bar\pi_{r})
\over \langle 1,2\rangle\langle 2,3\rangle\langle 3,4\rangle\langle 4,1\rangle}
{s\over t}\int \delta\left({(\xi_1-\xi_2)(\xi_3-\xi_4)\over(\xi_1-\xi_4)
(\xi_3-\xi_2)}+{s\over t}\right)\A^{J^A}_4
\eta(\tau)^4 \prod_{r=1}^4\rho_rd\nu_r{d\alpha_0
d\tau\over 2\pi\im\tau},$$

where we can take $\xi_r = \theta_3(2\hat\nu_r,2\tau)/
\theta_2(2\hat\nu_r,2\tau)$, $\hat\nu_r=\nu_r+i\alpha_0/4\pi$ and 
$\A_4^{J^A}$ is
given in (6.22) for a general compact Lie group.
We expect the gluon loop amplitude $\A_\floop$
to be related to the field theory loop amplitude
for gluons in $N=4$ Yang Mills theory coupled to $N=4$ conformal supergravity.
It is believed this field theory requires a dimension four gauge group to
avoid anomalies \refs{\RN}.
We hope that  a better understanding of that may follow from further analysis 
of the loop expressions computed in this paper.
\vfil\eject
\noindent{\bf Acknowledgements:}

We thank Edward Witten for discussions. 

LD thanks the Institute for Advanced Study at Princeton for its hospitality,
and was partially supported by the U.S. Department of Energy,
Grant No. DE-FG01-06ER06-01, Task A.
\vfill\eject
\appendix{A}{\hskip10pt Gauge Potentials}

An example of a potential on $S^2$, for which $A_z=A_\zb=0$, is 
$$\Aa_z^<=-{in\zb\over 1+z\zb},\qquad\Aat_z^<=0,$$
$$\Aa_z^>={in\over (1+z\zb)z},\qquad\Aat_z^>=0,$$
$$\Aa_\zb^<=0,\qquad\Aat_\zb^<=-{inz\over 1+z\zb},$$
$$\Aa_\zb^>=0,\qquad\Aat_\zb^>={in\over (1+z\zb)\zb}.$$
Then $\A^>_\mu-\A^<_\mu=-ig^{-1}\partial_\mu g$, $\quad\tilde\A^>_\mu-\tilde\A^<_\mu=-i\tilde g^{-1}\partial_\mu \tilde g$ for $\gg =z^{-n}$, $\ggt=\zb^{-n}$.

And an example of a potential on $T^2$, for which $A_z=A_\zb=0$, is
$$\Aa_z(z,\bar z) ={i\pi n\over {\rm Im}\tau}(z-\zb),\qquad
\Aat_z(z,\bar z) =0,$$
$$\Aa_\zb(z,\bar z)=0,\qquad\Aat_\zb(z,\bar z) =-{i\pi n\over {\rm Im}\tau}(z-\bar z),
$$
Then $$\eqalign{\A_\mu(z+a,\zb + \bar a) -\A_\mu(z,\zb) 
&=-ig^{-1}_a(\zb)\partial_\mu g_a(\zb), \cr
\tilde\A_\mu(z+a,\zb + \bar a)-\tilde\A_\mu(z,\zb)
&=-i\tilde g^{-1}_a(z)\partial_\mu \tilde g_a(z)\cr}$$ 
for 
$$\eqalign{
\gg_a(z) &= e^{-{\pi n (a- \bar a)\over {\rm Im}\tau} (z + {a\over 2})
+ i\pi n m_1n_1 + i \eta_a},\cr
\ggt_a(\zb) &= 
e^{{\pi n (a- \bar a)\over {\rm Im}\tau} (\zb + {\bar a\over 2})
- i\pi n m_1n_1 - i \bar\eta_a}.\cr}$$
\vfil\eject
\appendix{B}{\hskip10pt Twistors}

A Riemann surface world sheet, which we use in this paper,
corresponds to complex target space fields $Z^I$ that transform
under the conformal group $SU(2,2)$.
The spacetime metric is 
$\eta=\hbox{diag}(1,-1,-1,-1)$, and $\epsilon^{12}=-\epsilon^{21}
= -\epsilon_{12}=\epsilon_{21}=1$.
The coordinates are
$x=x^0-\bx\cdot\bsigma$, with $x^\dagger =x,$ and $\det x =x^\mu x_\mu$.
Then $$
x\mapsto (ax+b)(cx+d)^{-1},\qquad \pmatrix{a&b\cr c&d\cr}^\dagger
\pmatrix{0&-1_2\cr1_2&0\cr}\pmatrix{a&b\cr c&d\cr}=
\pmatrix{0&-1_2\cr1_2&0\cr},$$
where $\pmatrix{a&b\cr c&d\cr} \in SU(2,2)$.

The two 2-spinors $(\pi_a,\omega^{\dot a})$ define complex 2-planes
in complexified Minkowski space by
$\pi=x\omega$. Under a conformal transformation, the twistor
$$\pmatrix {\pi\cr\omega\cr} \mapsto\pmatrix{ a&b\cr c&d\cr}
\pmatrix{\pi\cr\omega\cr}.$$
Real Lorentz transformations  are given by
$x\mapsto axa^\dagger$, $\pi\mapsto a\pi$, $\omega\mapsto d\omega$
where $d=(a^\dagger)^{-1}$ and
$a, d\in\hbox{SL}(2,\Cop).$
For any vector $p^\mu$, we can write
$$
{\rm p}=p^\mu\sigma_\mu=\pmatrix{p^0-p^3&-p^1+ip^2\cr- p^1-ip^2&p^0+p^3\cr}
\equiv ({\rm p}_{a\da}),\qquad
{\rm p}^\dagger={\rm p},
$$
where $\epsilon^{ab}\epsilon^{\da\db}{\rm p}_{a\da}{\rm p}_{b\db}=2\det({\rm p})
=2p_\mu p^\mu\equiv 2p^2$,
and $\epsilon^{ab}\epsilon^{\da\db}{\rm p}_{a\da}\q_{b\db}=2p\cdot q.$
Then real Lorentz transformations are given by ${\rm p}\mapsto a{\rm p}
a^\dagger$,
$a\in\hbox{SL}(2,\Cop)$; and the complex Lorentz transformations are
${\rm p}\mapsto a{\rm p} b$, $a, b\in\hbox{SL}(2,\Cop).$

If $p^2=0$, we can write ${\rm p}=\lambda\bar\lambda^T$,
{\it i.e.} ${\rm p}_{a\da}=\lambda_a\bar\lambda_{\da}$.
Similarly, if $q^2=0$, $\q_{b\db}=\mu_b\bar\mu_{\db}$, then 
$2p\cdot q 
=\epsilon^{ab}\lambda_a\mu_b\,\epsilon^{\da\db}\bar\lambda_\da\bar\mu_\db
=\langle \lambda,\mu\rangle\, [\bar\lambda,\bar\mu],$
where
$\langle \lambda,\mu\rangle=\epsilon^{ab}\lambda_a\mu_b$
and $[\bar\lambda,\bar\mu]
=\epsilon^{\da\db}\bar\lambda_\da\bar\mu_\db.$
If
$\sum_{r=1}^n p_r=0$,  with ${p_r}_{a\da}={\pi_r}_a\bar\pi_{r\da}$,
then $\sum_{r=1}^n{\pi_r}_a\bar\pi_{r\da}=0$;
and $\sum_{r\ne s} \langle s r\rangle\bar\pi_{r\da}=0$ and
$\sum_{r\ne s} [s r]{\pi_r}_a=0,$
where $[r s]=[\bar\pi_r,\bar\pi_s]$, and $\langle r s\rangle
= \langle \pi_r,\pi_s\rangle$.
{\it E.g.}, for $n=3$,
${\pi_{1a}/ \pi_{2a}}=-{[2,3]/ [1,3]}$.

Under conformal transformations,
$$
\pmatrix{\lambda\cr\mu\cr}\mapsto\pmatrix{ a&b\cr c&d\cr}\pmatrix{\lambda\cr\mu\cr},\qquad
\pmatrix{\psi_\lambda\cr\psi_\mu\cr}\mapsto\pmatrix{a&b\cr c&d\cr}
\pmatrix{\psi_\lambda\cr\psi_\mu\cr},
$$
where $\psi_\lambda = \pmatrix{\psi^1\cr\psi^2\cr}$ and
$\psi_\mu= \pmatrix{\psi^3\cr\psi^4\cr}$.
\vfill\eject

\leftline{\sl Polarizations}

To compute the polarization dependence of the amplitudes as in \treemppfa,
we note that 
$s_{\dot a}$ and $\bar s_a$ are reference vectors that
scale as the momentum: $s_{\dot a}, \pi^a \,\sim u$ scale up and
$\bar s_a, \bar\pi^{\dot a} \,\sim u^{-1}$ scale down. The
polarization vector $\epsilon_{a\dot a\,r}$
does not scale, so $A_{-\,r}$ goes as $u^{-2}$ and
$A_{+\,r}$ as $u^2$. Hence the vertex operator
$(\pi^1)^{-2}\,
\left[ A_{+\,r} + ({\pi^1\over\lambda^1})^4 \,\psi^1\psi^2\psi^3\psi^4\,
A_{-\,r}\right]\,$
does not scale.
Then
$$\eqalignno{&\epsilon_1^-\cdot\epsilon_2^+ \epsilon_3^+\cdot p_1\,
+ \,\epsilon_2^+\cdot\epsilon_3^+ \epsilon_1^-\cdot p_2\,
+ \, \epsilon_3^+\cdot\epsilon_1^- \epsilon_2^+\cdot p_3\cr
&= A_{-1} A_{+2} A_{+3}\,\,\,
(\pi_1^b \bar s_2^a \bar s_{3b} \pi_{1a}\bar\pi^{3\dot b}\bar\pi_{2\dot b})
= A_{-1} A_{+2} A_{+3}\,\,\, {[23]^3\over [12][31]}\,.&\cr}$$
To derive this we use momentum conservation
$\sum_{r=1}^3 \pi_{ra}\bar\pi_{r\dot a} = 0$, and properties of the
Penrose spinors such as $\sum_{a=1}^2 \pi_r^a\pi_{ra}=\nobreak 0$,
where $\pi_{ra} = \epsilon_{ab}\pi_r^{b}$ and
$\pi_r^a = \epsilon^{ab}\pi_{rb}$.
In particular, consider
$\bar s_{3b} \, \sum_{r=1}^3 \pi_r^b\bar\pi_r^{\dot b}  = 0$ to
find $\bar s_{3b} \pi_1^b = {[23]/[12]}$, and similarly 
$\bar s_2^a \pi_{1a} = {[23]/[13]}$.

\vfil\eject
\appendix{C}{\hskip10pt Trace Calculations}

{\sl Bosonic Trace}

To calculate the $M$-point trace, as in \trinv\ let
$$
\Phi(\omega,k)=\tr\left(e^{q_0}e^{ik_1Z_0}e^{q_0}e^{ik_2Z_0}\ldots e^{q_0}e^{ik_dZ_0}u^{a_0}e^{i\omega_1 Z(\rho_1)}e^{i\omega_2 Z(\rho_2)}\ldots e^{i\omega_M Z(\rho_M)}w^{L_0}\right),
$$
so
$$
\Phi(0,k)=\prod_{i=1}^d\delta(k_i),
$$
because only the states with no non-zero modes contribute to this trace.
Let
$$
\Phi_n(\omega,k)=\tr\left(e^{q_0}e^{ik_1Z_0}e^{q_0}e^{ik_2Z_0}\ldots e^{q_0}e^{ik_dZ_0}u^{a_0}Z_ne^{i\omega_1 Z(\rho_1)}e^{i\omega_2 Z(\rho_2)}\ldots e^{i\omega_M Z(\rho_M)}w^{L_0}\right),
$$
so
$$\partial_{k_1}\Phi=iu\Phi_{1-d},\quad\ldots\quad
\partial_{k_{d-m}}\Phi=iu\Phi_{-m},\quad\ldots
\quad\partial_{k_d}\Phi=iu\Phi_0,
$$
using 
$$e^{q_0}Z_ne^{-q_0}=Z_{n+1},\qquad u^{a_0}Z_nu^{-a_0}=u^{-1}Z_n.$$

Then, using the cyclic property of trace,
$$
 \Phi_n=uw^n\Phi_{n-d}=u^{(n+m)/d}{w}^{ (n+d-m)(n+m)/2d}\Phi_{-m},\qquad 0\leq m<d,\qquad
 n+m \hbox{ a multiple of }d
 $$
and

$$\eqalign{
\partial_{\omega_j}\Phi = i\sum_{n=-\infty}^\infty\Phi_{n}\rho_j^{-n}
&=i\sum_{i=0}^{d-1}\sum_{n=-\infty}^\infty\Phi_{dn-i}\rho_j^{-dn+i}\cr
&=i\sum_{i=0}^{d-1}\Phi_{-i}\sum_{n=-\infty}^\infty u^n w^{\half n(d(n+1)-2i)}\rho_j^{-dn+i}\cr
&=i\sum_{i=0}^{d-1}\Phi_{-i}\sum_{n=-\infty}^\infty w^{\half (n-1)(dn-2i)}u^{n-i/d}\left({\rho_j\over w}\right)^{-dn+i}u^{i/d}\cr
&=i\sum_{i=0}^{d-1}\Phi_{-i}F^d_{d-i}(u^{-1/d}\rho_j,w)u^{i/d}\cr
&=\sum_{n=1}^{d}F^d_{n}(u^{-1/d}\rho_j,w)u^{-n/d}\partial_{k_n}\Phi\cr
}$$
where
$$
F^d_i(\rho_j,w)=\sum_{n=-\infty}^\infty w^{\half (n-1)(d(n-2)+2i)}\left({\rho_j\over w}\right)^{d(1-n)-i}
$$
So $\Phi$ depends on $k_i,\omega_j$, $1\leq i\leq d, 1\leq j\leq M,$ through 
$$u^{i/d}k_i+ \sum_{j=1}^MF^d_i(\hat\rho_j,w)\omega_j,\qquad\hbox{where } \hat\rho_j=u^{-1/d}\rho_j,$$

so that
$$\Phi(\omega,k)=u^{(d+1)/2}\prod_{i=1}^d\delta\left(u^{i/d}k_i+ \sum_{j=1}^MF^d_i(\hat\rho_j,w)\omega_j\right)
$$
and the $M$ point function

$$\eqalignno{
\tr\left(e^{dq_0}u^{a_0}e^{i\omega_1 Z(\rho_1)}e^{i\omega_2 Z(\rho_2)}\ldots e^{i\omega_M Z(\rho_M)}w^{L_0}\right)&\cr
&\hskip-167pt=\Phi(\omega,0)\cr
&\hskip-167pt=u^{(d+1)/2}\prod_{i=1}^d\delta\left(\sum_{j=1}^MF^d_i(\hat\rho_j,w)\omega_j\right)&(\hbox{C}.1)\cr
&\hskip-167pt=u^{(d+1)/2}
\left|\matrix{F^d_1(\hat\rho_1,w)&F^d_1(\hat\rho_2,w)&\ldots&F^d_1(\hat\rho_d,w)\cr
F^d_2(\hat\rho_1,w)&F^d_2(\hat\rho_2,w)&\ldots&F^d_2(\hat\rho_d,w)\cr
.&.&\ldots&.\cr.&.&\ldots&.\cr
F^d_d(\hat\rho_1,w)&F^d_d(\hat\rho_2,w)&\ldots&F^d_d(\hat\rho_d,w)\cr}\right|^{-1}
\prod_{m=1}^d\delta(\omega_m),&(\hbox{C}.2)\cr
}$$
where the last equality (C.2) holds provided that $M=d$.

{\sl Comparison with Bosonic Tree Amplitudes}

We can compare these results for corresponding results for tree amplitudes,
$$\bvac e^{dq_0} e^{i\omega_MZ(\rho_M)}\ldots e^{i\omega_0Z_0}\vac.$$
We start with 
$$\bvac e^{dq_0} e^{ik_dZ_{-d}}\ldots e^{ik_0Z_0)}\vac=\prod_{i=0}^d\delta(k_i).$$
Let
$$
\Phi = \bvac e^{dq_0} \prod_{i=0}^de^{ik_iZ_{-i} }\prod_{j=0}^Me^{i\omega_jZ(\rho_j)}\vac
$$
$$\eqalign{
\partial_{\omega_j}\Phi &= i\sum_{k=0}^d\bvac e^{dq_0} \prod_{i=0}^de^{ik_iZ_{-i} }\prod_{j=0}^Me^{i\omega_kZ(\rho_j)}Z_{-k}\vac\rho^k_j\cr
&=\sum_{i=0}^d\rho_j^i\partial_{k_i}\Phi.
}$$
These provide $M+1$ linear partial differential equations in the $M+d+2$ variables $k,\omega$, implying that $\Phi$ depends only on the $d+1$ variables
$$k_i+\sum_{j=0}^M\rho_j^i\omega_j$$
so that
$$
\Phi = \prod_{i=0}^d\delta\left(k_i+\sum_{j=0}^M\rho_j^i\omega_j\right).
$$
Thus
$$\eqalignno{
\bvac e^{dq_0} e^{i\omega_MZ(\rho_M)}\ldots e^{i\omega_0Z(\rho_0)}\vac
&=\prod_{i=0}^d\delta\left(\sum_{j=0}^M\rho_j^i\omega_j\right)\cr
&=\left|\matrix{1&1&\ldots&1\cr
\rho_0&\rho_1&\ldots&\rho_d\cr
.&.&\ldots&.\cr.&.&\ldots&.\cr
\rho_0^d&\rho_1^d&\ldots&\rho_d^d\cr}\right|^{-1}\prod_{j=0}^d\delta(\omega_j)\cr
&=\prod_{0\leq i<j\leq d}(\rho_i-\rho_j)^{-1}\prod_{j=0}^d\delta(\omega_j),&(\hbox{C}.3)\cr
}$$
provided  that $M=d$.

{\sl Fermionic Tree Amplitudes}

Starting from
$$\bvac e^{dq_0}Z_{-d}\ldots Z_{-1}Z_0\vac = 1,$$
so that
$$\bvac e^{dq_0}Z_{-{n_d}}\ldots Z_{-n_1}Z_{-n_0}\vac = \epsilon_{n_d\ldots n_1n_0},$$
where $\epsilon_{n_d\ldots n_1n_0}$ is totally antisymmetric and nonzero only if 
$(n_d,\ldots ,n_1,n_0)$ is a permutation of $(d,\ldots, 1,0)$ with $\epsilon_{d\ldots 10}=1$.
Then
$$\eqalignno{
\bvac e^{dq_0}c(\rho_d)\ldots c(\rho_1)c(\rho_0)\vac &= \sum_{n_j=-\infty}^\infty
\rho_d^{n_d}\ldots \rho_1^{n_1}\rho_0^{n_0}\bvac e^{dq_0}Z_{-{n_d}}\ldots Z_{-n_1}Z_{-n_0}\vac\cr
&=\sum_{n_j=0}^d \epsilon_{n_d\ldots n_1n_0}
\rho_d^{n_d}\ldots \rho_1^{n_1}\rho_0^{n_0}\cr
&=\left|\matrix{1&1&\ldots&1\cr
\rho_0&\rho_1&\ldots&\rho_d\cr
.&.&\ldots&.\cr.&.&\ldots&.\cr
\rho_0^d&\rho_1^d&\ldots&\rho_d^d\cr}\right|\cr
&=\prod_{0\leq i<j\leq d}(\rho_i-\rho_j),&(\hbox{C}.4)\cr
}$$
which can be compared with (C.2).

{\sl Fermionic Loop Amplitudes}

Consider 
$$\tr\left(e^{dq_0}u^{a_0}(-1)^{Na_0}Z(\rho_d)\ldots Z(\rho_2)Z(\rho_1)w^{L_0}\right),$$
noting that
$$\eqalign{\tr\left(e^{dq_0}u^{a_0}(-1)^{Na_0}Z_{-d}\ldots Z_{-2}Z_{-1}w^{L_0}\right)\hskip-50pt&\cr
&=u^{-d}(-1)^{Nd}\tr\left(e^{dq_0}Z_{-d}\ldots Z_{-2}Z_{-1}u^{a_0}(-1)^{Na_0}w^{L_0}\right)\cr
&=u^{-d}(-1)^{Nd}\langle 0|e^{dq_0}Z_{-d}\ldots Z_{-2}Z_{-1}u^{a_0}(-1)^{Na_0}w^{L_0}\vac\cr
&=u^{-d}(-1)^{Nd},\cr}$$
where $N$ is an integer, because again only the states with no non-zero modes contribute to this trace. 

If $\varphi(Z)$ denotes an arbitrary sum of products of the $Z_m$ each of length $d-1$,
$$
\Phi_n=\tr\left(e^{dq_0}u^{a_0}(-1)^{Na_0}Z_n\varphi(Z)w^{L_0}\right)=(-1)^{N+d-1}uw^n\Phi_{n-d}
$$
using the cyclic property of trace.

So, as for the bosonic traces,
$$
 \Phi_{dn-m}=\eta^n w^{\half n(d(n+1)-2m)}\Phi_{-m},\qquad 0\leq m<d
$$
where $\eta=(-1)^{N+d-1}u$ 
and, again,
$$\eqalign{
tr\left(e^{dq_0}u^{a_0}(-1)^{Na_0}Z(\rho_j)\varphi(Z)w^{L_0}\right)\hskip-80pt&\cr
&=\sum_{n=-\infty}^\infty\Phi_{n}\rho_j^{-n}
=\sum_{m=0}^{d-1}\sum_{n=-\infty}^\infty\Phi_{dn-m}\rho_j^{-dn+m}\cr
&=\sum_{m=0}^{d-1}\Phi_{-m}\sum_{n=-\infty}^\infty \epsilon^n u^n w^{\half n(d(n+1)-2m)}\rho_j^{-dn+m}\cr
&=\sum_{m=0}^{d-1}\Phi_{-m}\sum_{n=-\infty}^\infty \epsilon^n w^{\half (n-1)(dn-2m)}
 u^{n-m/d}\left({\rho_j\over w}\right)^{-dn+m} u^{m/d}\cr
&=\sum_{m=0}^{d-1} u^{m/d}\tr\left(e^{dq_0}u^{a_0}(-1)^{Na_0}Z_{-m}\varphi(Z)w^{L_0}\right)F^{d\epsilon}_{d-m}( u^{-1/d}\rho_j,w)\cr
}$$
where $\epsilon=(-1)^{N+d-1}$ and
$$
F^{d\epsilon}_{m}(\rho,w)=\sum_{n=-\infty}^\infty \epsilon^nw^{\half (n-1)(d(n-2)+2m)}\left({\rho\over w}\right)^{d(n-1)-m}
$$
Writing $\hat\rho_j=u^{-1/d}\rho_j$, it follows that
$$\eqalignno{
\tr\left(e^{dq_0}u^{a_0}(-1)^{Na_0}Z(\rho_d)\ldots Z(\rho_2)Z(\rho_1)w^{L_0}\right)\hskip-70truemm&\cr
&=\sum_{m_j=1}^d\tr\left(e^{dq_0}u^{a_0}(-1)^{Na_0}Z_{-m_d}\ldots Z_{-m_2}Z_{-m_1}w^{L_0} \right)\prod_{j=1}^d u^{(m_j-1)/d}
F^{d\epsilon}_{d-m_j+1}(\hat\rho_j,w)\cr
&= u^{-(d+1)/2}(-1)^{Nd}\sum_{m_j=1}^d
\prod_{j=1}^d\epsilon_{m_d\ldots m_2m_1}F^{d\epsilon}_{d-m_j+1}(\hat\rho_j,w)\cr
&= u^{-(d+1)/2}\sum_{m_j=1}^d
\prod_{j=1}^d\epsilon_{m_d\ldots m_2m_1}F^{d}_{d-m_j+1}(\hat\rho_j,w),&(\hbox{C}.5)\cr
}$$
provided that $N=0$ when $d$ is odd and $N=1$ when $d$ is even, because then 
$\epsilon=1$ and $F^{d1}_m(\rho,w)=F^d_m(\rho,w)$.
  {\it i.e.} a factor $(-1)^{a_0}$ is included when $d$ is even and no factor of $(-1)^{Na_0}$ is included if $d$ is odd. In particular, in this case, there is a precise cancellation between (C.2) and (C.5) in the case $M=d$. More generally, for $M\ne d$, the overall factors of $u$ cancel and the functions
 $F^{d}_m(\hat\rho,w)$ involved are the same.
 We shall use this factor in computing the trace for the twistor
string loop, and find agreement with the path integral derivation.
Since $[Q,a_0^i] = 0$, it still holds that only states in the
cohomology of $Q$ contribute to the trace following the discussion at the
end of section 3. 
 
\vfil\eject

\appendix{D}{\hskip10pt  Current Algebra Loop from Recurrence Relations}

We outline the derivation of the three and four-point current algebra
one-loop amplitude from the recurrence relations described in section 6.
For the three-point function, from \defreca and \defrecb,
$$\eqalignno {&\tr(J^a(\rho_1)J^b(\rho_2) J^c(\rho_3) w^{L_0})\cr
&= i f^{abc} (\rho_1\rho_2\rho_3)^{-1}\Big\{
\chi^{(2)}(\tau) \left (\tilde\Delta_1(\nu_{12},\tau)
+ \tilde\Delta_1(\nu_{23},\tau) + \tilde\Delta_1(\nu_{31},\tau)\right)\cr
&\hskip10pt + k\chi(\tau) \left(- \Delta_2^1(\nu_{23},\tau) +
( \tilde\Delta_1(\nu_{12},\tau) + \tilde\Delta_1(\nu_{31},\tau))
\Delta_2(\nu_{23},\tau)\right)\Big \}\cr
&\hskip 15pt +{d^{abc}\over2\rho_1\rho_2\rho_3}\chi^{(3)}(\tau)\cr
&= i f^{abc} (\rho_1\rho_2\rho_3)^{-1}\cr
&\hskip10pt \cdot \Big\{
\left ( \chi^{(2)}(\tau) + k \chi(\tau) \Delta_2(\nu_{23},\tau)\right)
\left (\tilde\Delta_1(\nu_{12},\tau)
+ \tilde\Delta_1(\nu_{23},\tau) + \tilde\Delta_1(\nu_{31},\tau)\right)\cr
&\hskip40pt -{1\over 4\pi i}  
k\chi(\tau) \partial_\nu\Delta_2(\nu_{23},\tau)
\Big\}+{d^{abc}\over2\rho_1\rho_2\rho_3}\chi^{(3)}(\tau)\cr}$$

\noindent where $\nu_{ij} = \nu_j-\nu_i$, and from \defreca\ 
we encounter propagators in addition to those in section 6:
$$\eqalignno{
\tilde\Delta_1(\nu,\tau)&=\Delta_1(\nu,\tau) + {1\over 2}\cr
\Delta_2(\nu,\tau)&
= {1\over 2\pi i} \partial_\nu \Delta_1(\nu,\tau)
= {1\over 4\pi^2} \left({\theta_1'(\nu)\over\theta_1(\nu)}\right)'
= \chi_F^2(\nu,\tau) + {1\over 4\pi^2} {\theta''_3(0,\tau)\over
\theta_3(0,\tau)}\cr
\Delta_1^1(\nu,\tau)&=\sum_{m\ne 0}{e^{2\pi im\nu} w^m\over (1-w^m)^2}
= \sum_{m\ne 0}{e^{2\pi im\nu}\over (1-w^m)^2} - \Delta_1(\nu,\tau)\cr
&= \half \Delta_2(\nu,\tau) - \half (\tilde \Delta_1(\nu,\tau))^2 + {1\over 12}
-{1\over 24\pi^2}{\theta'''_1(0,\tau)\over\theta'_1(0,\tau)}\cr
\Delta_2^1(\nu,\tau)&=\sum_{m\ne 0}{me^{2\pi im\nu} w^m \over (1-w^m)^2}
= {1\over 2\pi i} \partial_\tau \Delta_1(\nu,\tau)
= {1\over 2\pi i} \partial_\nu \Delta_1^1(\nu,\tau)\cr
&= {i\over 16\pi^3} {\theta'_1(\nu,\tau)\theta''_1(\nu,\tau)
-\theta'''_1(\nu,\tau)\theta_1(\nu,\tau)\over\theta_1^2(\nu,\tau)}
= {1\over 4\pi i} \partial_\nu\Delta_2(\nu,\tau) - \tilde\Delta_1(\nu,\tau)
\Delta_2(\nu,\tau)\,. \cr}$$

\vfill\eject
Let
$$\tilde{\cal P}(\nu) = -4\pi^2\chi_F^2(\nu,\tau) - \theta_3^{-1}(0,\tau)
\theta_3''(0,\tau) - 4\pi^2 {\chi^{(2)}(\tau)\over \chi(\tau)}\,,$$
which depends on the partition function $\chi(\tau)$ and
the propagator $\chi^{(2)}(\tau)$ of a given representation.
It is related to the standard Weierstrass P function ${\cal P}(\nu,\tau)$
by an additive function of $\tau$. Then
$$\eqalignno {\tr&(J^a(\rho_1)J^b(\rho_2) J^c(\rho_3) w^{L_0})\cr
&= i f^{abc} (\rho_1\rho_2\rho_3)^{-1} k \chi(\tau) (-{i\over 16\pi^3})
\Big\{2 \tilde{\cal P}_{23}  \left(\zeta^{23}+\zeta^{31} +\zeta^{12}\right)
+ \tilde{\cal P}'_{23}\Big\}+ {d^{abc}\over2\rho_1\rho_2\rho_3}\chi^{(3)}(\tau).\cr}$$
Using the Weierstrass addition formulae, which hold for both ${\cal P}(\nu,\tau)$
and $\tilde{\cal P}(\nu,\tau)$:
$$\eqalignno {\zeta^{23} &= \zeta^{13} + \zeta^{21} + \half
{{\cal P}'_{13} - {\cal P}'_{21}\over {\cal P}_{13} - {\cal P}_{21}}\cr
{\cal P}'_{23} &=  {{\cal P}_{23} [{\cal P}'_{21}-{\cal P}'_{13}]
+ {\cal P}_{21} {\cal P}'_{13} - {\cal P}'_{21} {\cal P}_{13}\over
{\cal P}_{13} - {\cal P}_{21}}\,,\cr}$$
we can write
$$\eqalignno {\tr&(J^a(\rho_1)J^b(\rho_2) J^c(\rho_3) w^{L_0})\cr
&= i f^{abc} (\rho_1\rho_2\rho_3)^{-1} k \chi(\tau) (-{i\over 16\pi^3})
\Big\{ \tilde{\cal P}_{23} {\tilde{\cal P}'_{13}
- \tilde{\cal P}'_{21}\over {\cal P}_{13} - \tilde{\cal P}_{21}}
+ \tilde{\cal P}'_{23}\Big\}+ {d^{abc}\over2\rho_1\rho_2\rho_3}\chi^{(3)}(\tau)\cr
&= i f^{abc} (\rho_1\rho_2\rho_3)^{-1} k \chi(\tau) (-{i\over 16\pi^3})
\Big\{ {\tilde{\cal P}_{21} \tilde{\cal P}'_{13} - \tilde{\cal P}'_{21} 
\tilde{\cal P}_{13}\over
\tilde{\cal P}_{13} - \tilde{\cal P}_{21}}\Big \} 
+ {d^{abc}\over2\rho_1\rho_2\rho_3}\chi^{(3)}(\tau)\cr
&= i f^{abc} (\rho_1\rho_2\rho_3)^{-1} k \chi(\tau)\Big\{
\chi_F^{21}\chi_F^{32}\chi_F^{13}\,
- {i\over 2\pi} \left(\zeta^{21}
+ \zeta^{32} + \zeta^{13}\right)
f(\tau)\Big\}\cr
&\hskip15pt + (\rho_1\rho_2\rho_3)^{-1} 
{d^{abc}\over2\rho_1\rho_2\rho_3}\chi^{(3)}(\tau)\cr}$$
where the last line follows from standard theta function addition formulae
\refs{\Mumford}. 
\vfil\eject
In a similar way, we compute from the recurrence relation the four-point loop as 
$$\eqalignno{\tr& (J^a(\rho_1)J^b(\rho_2)J^c(\rho_3)J^d(\rho_4) w^{L_0})\cr
&= (\rho_1\rho_2\rho_3\rho_4)^{-1}\Big\{
\delta^{ab}\delta^{cd}
\Big( \left[{\chi^{(2)} (\tau)\over\chi(\tau)} +
k\Delta_2(\nu_{12},\tau)\right] \left [ \chi^{(2)}(\tau)
+ k \chi(\tau) \Delta_2(\nu_{34},\tau) \right]\cr
&\hskip110pt - \chi^{-1}(\tau) (\chi^{(2)}(\tau))^2\Big)\cr
&\hskip70pt + \delta^{ac}\delta^{bd}
\Big( \left[{\chi^{(2)} (\tau)\over\chi(\tau)} +k\Delta_2(\nu_{13},\tau)
\right] \left [ \chi^{(2)}(\tau)
+ k \chi(\tau) \Delta_2(\nu_{24},\tau) \right]\cr
&\hskip110pt - \chi^{-1}(\tau) (\chi^{(2)}(\tau))^2\Big)\cr
&\hskip70pt + \delta^{ad}\delta^{bc}
\Big( \left[{\chi^{(2)} (\tau)\over\chi(\tau)}
+ k\Delta_2(\nu_{14},\tau)\right] \left [ \chi^{(2)}(\tau)
+ k \chi(\tau) \Delta_2(\nu_{23},\tau) \right]\cr
&\hskip110pt - \chi^{-1}(\tau) (\chi^{(2)}(\tau))^2\Big)\cr
&\hskip70pt + \tr (J^a_0 J^b_0 J^c_0J^d_0 w^{L_0})\cr
&\hskip70pt + f^{ab}_{\hskip10pt e}f^{cde} \Big[-\half\chi^{(2)}(\tau)
\Delta_1(\nu_{34},\tau)\cr
&\hskip90pt + \left(\Delta_1^1(\nu_{23},\tau)
-\tilde\Delta_1^1(\nu_{24},\tau)\right) \left(\chi^{(2)}(\tau)
+ k\chi(\tau)\Delta_2(\nu_{34},\tau)\right)\cr
&\hskip90pt - \tilde \Delta_1(\nu_{12},\tau) \,
[\chi^{(2)}(\tau) (\tilde\Delta_1(\nu_{23},\tau)
+ \tilde\Delta_1(\nu_{34},\tau) + \tilde\Delta_1(\nu_{42},\tau))\cr
&\hskip90pt + k\chi(\tau) (- \Delta_2^1(\nu_{34},\tau) +
( \tilde\Delta_1(\nu_{23},\tau) + \tilde\Delta_1(\nu_{42},\tau))
\Delta_2(\nu_{34},\tau))]\,\Big]\cr
&\hskip70pt + f^{ac}_{\hskip10pt e}f^{bde}
\Big[ - \chi^{(2)}(\tau) \Delta_1^1(\nu_{34},\tau) + k \chi(\tau)
\Delta_3(\nu_{34},\tau)\cr
&\hskip90pt -\chi^{(2)}(\tau)\tilde\Delta_1(\nu_{34},\tau)
\Delta_1(\nu_{24},\tau)
+ k \chi(\tau) \Delta_2^1(\nu_{34},\tau) \Delta_1(\nu_{24},\tau)\cr
&\hskip90pt + k \chi(\tau) \Delta_2^1(\nu_{34},\tau)\cr
&\hskip90pt +\tilde\Delta_1(\nu_{13},\tau) \,
[\chi^{(2)}(\tau) (\tilde\Delta_1(\nu_{23},\tau)
+ \tilde\Delta_1(\nu_{34},\tau) + \tilde\Delta_1(\nu_{42},\tau))\cr
&\hskip90pt + k\chi(\tau) (- \Delta_2^1(\nu_{34},\tau) +
( \tilde\Delta_1(\nu_{23},\tau) + \tilde\Delta_1(\nu_{42},\tau))
\Delta_2(\nu_{34},\tau))]\,\Big]\cr
&\hskip70pt + f^{ad}_{\hskip10pt e}f^{bce} \Big[ - \chi^{(2)}(\tau) \Delta_1^1(\nu_{34},\tau)
+ k \chi(\tau) \Delta_3(\nu_{34},\tau)\cr
&\hskip90pt+ \chi^{(2)}(\tau)
\tilde\Delta_1(\nu_{34},\tau)\Delta_1(\nu_{23},\tau)
- k \chi(\tau) \Delta_2^1(\nu_{34},\tau)
\Delta_1(\nu_{23},\tau)\cr
&\hskip90pt -\tilde\Delta_1(\nu_{14},\tau) \,
[\chi^{(2)}(\tau) (\tilde\Delta_1(\nu_{23},\tau)
+ \tilde\Delta_1(\nu_{34},\tau) + \tilde\Delta_1(\nu_{42},\tau))\cr
&\hskip90pt + k\chi(\tau) (- \Delta_2^1(\nu_{34},\tau) +
( \tilde\Delta_1(\nu_{23},\tau) + \tilde\Delta_1(\nu_{42},\tau))
\Delta_2(\nu_{34},\tau))]\,\Big]\cr
&\hskip70pt
 + {i\over 2}\chi^{(3)}(\tau)\,
\Big( f^{ab}_{\hskip10pt e}d^{ecd}\Delta_1(\nu_{12})
+ f^{ac}_{\hskip10pt e}d^{bed}\Delta_1(\nu_{13})
+ f^{ad}_{\hskip10pt e}d^{bce}\Delta_1(\nu_{14})\cr
&\hskip100pt + f^{bc}_{\hskip10pt e}d^{aed}\Delta_1(\nu_{23})
+ f^{bd}_{\hskip10pt e}d^{ace}\Delta_1(\nu_{24})
+ f^{cd}_{\hskip10pt e}d^{abe}\Delta_1(\nu_{34})\Big)
\Big\},\cr}$$
\vfill\eject

where an additional propagator occurs,
$$\eqalignno{
\Delta_3(\nu,\tau)&=\sum_{m\ne 0}{m e^{2\pi im\nu} w^{2m} \over (1-w^m)^3}
= -\half\Delta_2^1(\nu,\tau)
+ {1\over 4\pi i} \partial_\tau\Delta_1^1(\nu,\tau)\cr
&= -\half\Delta_2^1(\nu,\tau) -\half \tilde\Delta_1(\nu,\tau)\Delta_2^1(\nu,\tau)
+ {1\over 8\pi i} \partial_\nu\Delta_2^1(\nu,\tau) + {i\over 96\pi^3}
\partial_\tau \left( {\theta'''_1(0,\tau)\over\theta'_1(0,\tau)}\right).
\cr}$$
This form of the four-point current algebra loop can be expressed in
terms of the Weierstrass $\P$ functions: 
$$\eqalignno{\tr& (J^a(\rho_1)J^b(\rho_2)J^c(\rho_3)J^d(\rho_4) w^{L_0})\rho_1\rho_2\rho_3\rho_4\cr
&=\Big\{
\delta^{ab}\delta^{cd}\, \left ({1\over 16\pi^2} k^2 \chi(\tau)\,
\tilde{\cal P}_{12}\tilde{\cal P}_{34}
- \chi^{-1}(\tau)(\chi^{(2)}(\tau))^2\right)\cr
&\hskip40pt + \delta^{ac}\delta^{bd}
\, \left( {1\over 16\pi^2} k^2 \chi(\tau)\,
\tilde{\cal P}_{31}\tilde{\cal P}_{24} -\chi^{-1}(\tau)(\chi^{(2)}(\tau))^2\right)\cr
&\hskip40pt + \delta^{ad}\delta^{bc}
\, \left({1\over 16\pi^2} k^2 \chi(\tau)\,
\tilde{\cal P}_{14}\tilde{\cal P}_{32} -  \chi^{-1}(\tau)(\chi^{(2)}(\tau))^2\right)\cr
&\hskip40pt + \tr (J^a_0 J^b_0 J^c_0J^d_0 w^{L_0})\cr
&\hskip40pt + f^{ab}_{\hskip10pt e}f^{cde} \big [
{1\over 64\pi^4} k\chi(\tau) \Big \{
{{\cal P}_{32} {\cal P}'_{24} 
- {\cal P}'_{32} {\cal P}_{24} \over {\cal P}_{24}  - {\cal P}_{32} }\,
\left (- {{\cal P}'_{24}  - {\cal P}'_{32} \over
{\cal P}_{24}  - {\cal P}_{32} } + {{\cal P}'_{24}  + {\cal P}'_{14}\over
{\cal P}_{24}  - {\cal P}_{14}}\right)\cr
&\hskip 70pt + 4 \tilde{\cal P}_{34} \tilde{\cal P}_{32} \Big\}\cr
&\hskip45pt 
+  k\chi(\tau) \,\,\Big\{- \left ({\chi^{(2)}(\tau)\over k\chi(\tau)}
+ {1\over 4\pi^2} {\theta''_3(0,\tau)\over\theta_3(0,\tau)}
\right)^2
+ {1\over 48} \theta_2^4(0,\tau)\theta_4^4(0,\tau)
+ {1\over 3} {\chi^{(2)}(\tau)\over k\chi(\tau)}\cr
&\hskip60pt
- {1\over 6} \left ({\chi^{(2)}(\tau)\over k\chi(\tau)}
+ {1\over 4\pi^2} {\theta''_3(0,\tau)\over\theta_3(0,\tau)}\right)
\left (\theta_2^4(0,\tau) - \theta_4^4(0,\tau)\right )
\Big\}\big]\cr
&\hskip40pt + f^{ac}_{\hskip10pt e}f^{bde} \big [
{1\over 64\pi^4} k   \chi(\tau)\,\{ {{\cal P}_{32} {\cal P}'_{24} 
- {\cal P}'_{32} {\cal P}_{24} \over {\cal P}_{24}  - {\cal P}_{32} } \,
\left({{\cal P}'_{14} - {\cal P}'_{31}\over
{\cal P}_{14} - {\cal P}_{31}} + {{\cal P}'_{24}  - {\cal P}'_{32} \over
{\cal P}_{24}  - {\cal P}_{32} }\right )\cr
&\hskip60pt 
- 4 \tilde{\cal P}_{34} (\tilde{\cal P}_{24}  + \tilde{\cal P}_{32}  )\} \cr
&\hskip45pt
+ k\chi(\tau) \,\,\Big\{2 \left ({\chi^{(2)}(\tau)\over k\chi(\tau)}
+ {1\over 4\pi^2} {\theta''_3(0,\tau)\over\theta_3(0,\tau)}
\right)^2
 - {1\over 24} \theta_2^4(0,\tau)\theta_4^4(0,\tau)
-{1\over 6} {\chi^{(2)}(\tau)\over k\chi(\tau)}\cr
&\hskip60pt
+ {1\over 3} \left ({\chi^{(2)}(\tau)\over k\chi(\tau)}
+ {1\over 4\pi^2} {\theta''_3(0,\tau)\over\theta_3(0,\tau)}\right)
\left (\theta_2^4(0,\tau) - \theta_4^4(0,\tau)\right )
\Big\}\Big]\cr
& \hskip45pt + {i\over 2}\chi^{(3)}(\tau)\,
\Big( f^{ab}_{\hskip10pt e}d^{ecd}\Delta_1(\nu_{12})
+ f^{ac}_{\hskip10pt e}d^{bed}\Delta_1(\nu_{13})
+ f^{ad}_{\hskip10pt e}d^{bce}\Delta_1(\nu_{14})\cr
&\hskip80pt + f^{bc}_{\hskip10pt e}d^{aed}\Delta_1(\nu_{23})
+ f^{bd}_{\hskip10pt e}d^{ace}\Delta_1(\nu_{24})
+ f^{cd}_{\hskip10pt e}d^{abe}\Delta_1(\nu_{34})\Big)
\Big\}.\cr}$$
\vfill\eject
\noindent We introduce traces over the group matrices to express the
combinations of structure constants and d-symbols that
occur in the four-point loop. For $\sigma^{abcd} \equiv \tr(T^aT^bT^cT^d)$,
$$\eqalignno{\sigma^{abcd} - \sigma^{adcb}
+ \sigma^{acdb} - \sigma^{abdc} &= i f^{cd}_{\hskip10pt e}d^{abe},\cr
-\sigma^{abcd} - \sigma^{adcb}
+ \sigma^{acdb} + \sigma^{abdc} &= 2k\,\,f^{ab}_{\hskip 10pt e}f^{cde}.\cr}$$
The symmetrization of the trace
$\tr (J^a_0 J^b_0 J^c_0J^d_0 w^{L_0})$ over permutations of $a,b,c,d$ is
$$\eqalignno{&\tr(J_0^aJ_0^bJ_0^cJ_0^d w^{L_0})\cr& =
\left(\tr(J_0^aJ_0^bJ_0^cJ_0^d w^{L_0})\right)_{\bf S}\cr
&+\hskip5pt{i\over 4} \left(
f^{cd}_{\hskip10pt e}d^{abe} + f^{bd}_{\hskip10pt e}d^{ace} + f^{bc}_{\hskip10pt e}d^{ade}\right) \chi^{(3)}(\tau)
+{1\over 6} \left(f^{bc}_{\hskip10pt e}f^{ade} - f^{cd}_{\hskip10pt e}f^{abe}\right) \chi^{(2)}(\tau).\cr}$$
Then the four-point loop can be written in the symmetric form \fourg.

\listrefs

\end